\documentclass[journal=jpcbfk,manuscript=article]{achemso}
\usepackage{longtable}
\usepackage{multirow}
\usepackage{amsmath}
\usepackage[usenames,dvipsnames]{color}
\usepackage{soul}
\usepackage{threeparttable}
\usepackage{graphicx}
\usepackage{amsmath,amssymb}
\usepackage{caption}
\usepackage{color}
\usepackage{dcolumn}
\usepackage{bm}
\usepackage{float}
\usepackage{subcaption}
\usepackage{textcomp}
\usepackage{url}
\usepackage[normalem]{ulem}
\usepackage{booktabs}

\usepackage{microtype}

\author{Kai T\"opfer} \affiliation[University of Basel]{Department of
  Chemistry, University of Basel, Klingelbergstrasse 80 , CH-4056
  Basel, Switzerland}

\author{Debasish Koner} \affiliation[University of Basel]{Department
  of Chemistry, University of Basel, Klingelbergstrasse 80 , CH-4056
  Basel, Switzerland} \altaffiliation{Department of Chemistry, Indian
  Institute of Technology Hyderabad, Sangareddy, Telangana 502285,
  India}

\author{Shyamsunder Erramilli} \affiliation[Boston University]{
  Department of Chemistry and the Photonics Center, Boston University,
  8 St Mary's St, MA 02215, Boston}

\author{Lawrence D. Ziegler} \affiliation[Boston University]{
  Department of Chemistry and the Photonics Center, Boston University,
  8 St Mary's St, MA 02215, Boston}

\author{Markus Meuwly} \affiliation[University of Basel]{Department of
  Chemistry, University of Basel, Klingelbergstrasse 80 , CH-4056
  Basel, Switzerland} \altaffiliation{Department of Chemistry, Brown
  University, Providence, RI 02912, USA} \email{m.meuwly@unibas.ch}

\title{Molecular-Level Understanding of the Ro-vibrational Spectra of
  N$_2$O in Gaseous, Supercritical and Liquid SF$_6$ and Xe}

\begin{document}

\section{Abstract}
The transition between the gas-, supercritical-, and liquid-phase
behaviour is a fascinating topic which still lacks molecular-level
understanding. Recent ultrafast two-dimensional infrared spectroscopy
experiments suggested that the vibrational spectroscopy of N$_2$O
embedded in xenon and SF$_6$ as solvents provides an avenue to
characterize the transitions between different phases as the
concentration (or density) of the solvent increases. The present work
demonstrates that classical molecular dynamics simulations together
with accurate interaction potentials allows to (semi-)quantitatively
describe the transition in rotational vibrational infrared spectra
from the P-/R-branch lineshape for the stretch vibrations of N$_2$O at
low solvent densities to the Q-branch-like lineshapes at high
densities. The results are interpreted within the classical theory of
rigid-body rotation in more/less constraining environments at high/low
solvent densities or based on phenomenological models for the
orientational relaxation of rotational motion. It is concluded that
classical MD simulations provide a powerful approach to characterize
and interpret the ultrafast motion of solutes in low to high density
solvents at a molecular level.

\section{Introduction}
Solvent-solute interactions and the coupled dynamics between solute
molecules embedded in a solvent are central for understanding
processes ranging from rotational and vibrational energy relaxation to
chemical reactivity in solution.\cite{stratt:1996,taylan:2022}
Vibrational spectroscopy is a particularly suitable technique to
follow the structural dynamics of "hot" solutes in their electronic
ground state interacting with a solvent environment. Experimentally,
1d- and 2d-infrared spectroscopies provide measures of the strength
and time scale of solvent-solute interactions that couple to the
resonant rovibrational excitation.\cite{hamm:2011} These effects can
also be probed directly by molecular dynamics (MD)
simulations.\cite{MM.rev:2022} The comparison of computations and
experiments is a sensitive test for the quality of the intermolecular
interaction potentials and provide a molecular-level understanding of
energy transfer mechanisms. Furthermore, the entire structural
dynamics at molecular-level detail is contained in the time dependent
motion of all solution species involved from the MD trajectories and
available for further analysis.\\

The understanding of solute equilibration in solvents including high
density gases and supercritical fluids (SCFs) is also important from a
more practical perspective. For example, controlling the outcome of
combustion reactions which are often performed in the supercritical
regime requires knowledge of energy transfer rates and mechanisms in
high temperature and pressure solutions.\cite{bares:2017,hessel:2022}
For reactions, it has been recognized that characterizing and
eventually tuning the physico-chemical properties of the solvent can
be as important as determining the best catalyst.\cite{hessel:2022}
Hence, gaining molecular-level insight into the properties of
solute-solvent dynamics is central not only for understanding
fundamental solution interactions but also for industrial processes
where solvents such as ionic, supercritical, or eutectic liquids are
employed. On the other hand, the possibility to specifically
manipulate dynamical properties in solvent system have been already
successfully exploited in a wide range of
applications.\cite{kajimoto:1999,skerget:2014} \\

Linear and nonlinear optical and in particular infrared (IR)
spectroscopy is a powerful means to characterize the structural and
dynamical properties of condensed-phase systems. As an example, recent
spectroscopic experiments in the Terahertz and infrared region
combined with MD simulations provided an atomistically resolved
picture of the structural properties of an eutectic
mixture.\cite{eutectic:2022} More broadly speaking, the structural and
dynamical properties of solutions can and have been studied at a
molecular level using two-dimensional IR (2DIR) spectroscopy through
vibrational energy
transfer.\cite{kuo07,bian2011ionclustering,bian2011nonresonant,bian2012ion,bian2013cation,chen2012ultrafast,Chen2014b,Chen2014a,Fernandez-Teran2020a}
The energy transfer rate for exchange of vibrational energy is
expected to follow a 6th-power, distance dependent
law,\cite{Chen2014a} similar to NOESY in NMR
spectroscopy,\cite{Schirmer1970,Bell1970} or F\"orster energy transfer
between electronic chromophores.\cite{Stryer1967} As an example, the
presence of cross-peaks in a 2D spectrum has been directly related to
the formation of aggregated structures.\cite{bian2011ionclustering}
However, interpretation and more in-depth understanding of the solvent
structure requires additional information that can be obtained for
example from MD simulations. \\

Recent 2DIR experiments on gas and supercritical phase solutions have
shown how rotational energy returns to equilibrium following
rovibrational excitation.\cite{ziegler:2018,ziegler:2019,ziegler:2022}
The dynamics of N$_2$O as the spectroscopic probe surrounded by xenon
atoms and SF$_6$ molecules has provided valuable information about
density-dependent change in the solvent structure and energy transfer
dynamics as the fluid approaches and passes through the near-critical
point density region. The 1D and 2D spectra of the N$_2$O asymmetric
stretch $(\nu_3)$ in Xe and SF$_6$ exhibit a significant dependence on
solvent density.\cite{ziegler:2019} At low density, corresponding to
SF$_6$ and Xe in the gas phase, the FTIR (1D) band shape of the
asymmetric stretch vibration is that of gas-phase N$_2$O with clearly
resolved P- and R-band structure whereas at high solvent density of
the liquid Xe or SF$_6$ solvent only a Q-branch-like absorption
feature peaked at the pure vibrational transition frequency was
observed. Because the solvent density $\rho$ can be changed in a
continuous fashion between gas and liquid densities, along a
near-critical isotherm $(T \sim 1.01 T_c)$ it is also possible to probe
the intermediate, super-critical regime of the solvent. 2DIR spectral
shapes as a function of waiting time and solvent density exhibit
perfectly anti-correlated features that report on the $J-$scrambling
or rotational energy relaxation
rates.\cite{ziegler:2018,ziegler:2019,ziegler:2022} \\

Given the molecular-level detail provided by quantitative MD
simulations, the present work focuses on changes in the IR
spectroscopy of N$_2$O embedded in SF$_6$ and xenon and their
interpretation at a structural level. For this, an accurate
kernel-based representation of high-level electronic structure
calculations for the spectroscopic reporter (N$_2$O) is combined with
state-of-the art treatment of electrostatic interactions employing the
minimally distributed charge model (MDCM)
\cite{MM.dcm:2014,MM.mdcm:2017,devereux:2020mdcm} which captures
multipolar interactions. Extensive MD simulations for densities
corresponding to those used in the recent 1D and 2D experiments were
carried out.\\

The current work is structured as follows. First, the Methods are
presented. Next, the quality of the interaction potentials is
discussed. Then, the density-dependent FTIR spectra are compared with
experiments, the vibrational frequency fluctuation correlation
functions are characterized and the organization of the solvent is
analyzed for xenon and SF$_6$ as the solvents. Finally, the findings
are discussed in a broader context.\\

\section{Methods}

\subsection{Potential Energy Surfaces}
The intramolecular potential energy surface (PES) of N$_2$O in its
electronic ground state ($^1$A$'$) is provided by a machine-learned
representation using the reproducing kernel Hilbert space (RKHS)
method.\cite{MM.rkhs:2017,koner:2020} Reference energies at the
CCSD(T)-F12/aug-cc-pVQZ level of theory were determined on a grid of
Jacobi coordinates $(R,r,\theta)$ with $r$ the N-N separation, $R$ the
distance between the center of mass of the diatom and the oxygen atom,
and $\theta$ the angle between the two distance vectors. The grid in
$r$ contained 15 points between 1.6\,a$_0$ and 3.1\,a$_0$, 15 points
between 1.8\,a$_0$ and\,4.75 a$_0$ for $R$, and a 10-point
Gauss-Legendre quadrature between $0^\circ$ and $90^\circ$ for
$\theta$. All calculations were carried out using the MOLPRO
package.\cite{werner:2020}\\

The total energy is represented by a 3D kernel as
\begin{equation}
     K({\bf{x}}, {\bf{x}}') =  k^{[2,6]}(R,R') k^{[2,6]}(r,r')  k^{[2]}(z,z')
\end{equation}
and the total PES is represented as $V({\bf{x}}) = \sum_{i=1}^N c_i
K({\bf{x}}, {\bf{x}}')$. For more details the reader is referred to
the literature.\cite{MM.rkhs:2017} Reciprocal power decay kernels
\begin{equation}
k^{2,6}(x,x') = \frac{1}{14}\frac{1}{x_>^7}    -\frac{1}{18}\frac{x_<}{x_>^8} 
\end{equation}
were used for the radial dimensions ($R$ and $r$). Here, $x_>$ and
$x_<$ are the larger and smaller values of $x$ and $x'$,
respectively. For large separations such kernels approach zero
according to $\propto \frac{1}{x^n}$ (here $n=6$) which gives the
correct long-range behavior for neutral atom-diatom type
interactions. For the angular coordinate a Taylor spline kernel
$k^{[2]}(z,z') = 1 + z_< z_> + 2z_<^2 z_> - \frac{2}{3} z_<^3$ is used
where $z = \frac{1 - {\rm cos} \theta}{2}$ and $z_>$ and $z_<$ are
again the larger and smaller values of $z$ and $z'$, defined similarly
to $x_>$ and $x_>$.\\

Intramolecular and intermolecular force field parameters for SF$_6$
are those from the work of Samios {\it et al.}.\cite{samios:2010}
Intermolecular interactions are based on Lennard-Jones potentials only
and the parameters are optimized such that MD simulations of pure
SF$_6$ reproduce the experimentally observed $PVT$ state points for
liquid and gas SF$_6$, as well as the states of liquid–vapor
coexistence below and supercritical fluid above the critical
temperature $T_c$, respectively. For xenon the parametrization for a
Lennard-Jones potential from Aziz {\it et al.} was used that
reproduces dilute gas macroscopic properties such as virial
coefficient, viscosity and thermal conductivity over a wide
temperature range but not specifically supercritical fluid
properties.\cite{aziz:1986} As discussed further below, the critical
concentration for xenon as determined from simulations (see Figure
S1) was found to be 5.19\,M compared with the
experimentally reported value\cite{ambrose:1987,haynes:2014crc} of
8.45\,M and variations for the critical temperature $T_c$ for xenon
compared with experiments can also be anticipated. In comparison, the
critical concentration of $\sim 5$\,M for SF$_6$ from the present work
(see below) matches the experimental value of
$5.06$\,M\cite{ambrose:1987,haynes:2014crc} and is close to 5.12\,M
(obtained from the reported critical density of 0.74 g/ml) from the
parametrization study.\cite{samios:2010}\\

Electrostatic interactions are computed based on a minimally
distributed charge model that correctly describes higher-order
multipole moments of a
molecule.\cite{MM.dcm:2014,MM.mdcm:2017,devereux:2020mdcm} For
parametrization, a reference electrostatic potential (ESP) of N$_2$O
in the linear equilibrium conformation is computed at the
CCSD/aug-cc-pVTZ level using the Gaussian program
package.\cite{gaussian16} The optimized MDCM fit reproduces the ESP
with a RMSE of $0.31$\,kcal/mol. For SF$_6$ in its octahedral
equilibrium conformation the ESP is computed at the MP2/aug-cc-pVTZ
level of theory using the Gaussian program and the RMSE of the fitted
ESP from MDCM is $0.11$\,kcal/mol. Recently,\cite{devereux:2020mdcm}
non-iterative polarization was also included in MDCM, and this is also
used here for N$_2$O, SF$_6$ and Xe. The polarizability of linear
N$_2$O computed at the CCSD/aug-cc-pVTZ level is $2.85$\,\AA$^3$ (with
each atom contributing $\sim 0.95$\,\AA$^3$ per atom), compared with
$2.998$\,\AA$^3$ from experiment.\cite{olney:1997} For Xe at the
CCSD/aug-cc-pVTZ the computed value of $2.96$\,\AA$^3$ compares with
$4.005$\,\AA$^3$ from experiment.\cite{olney:1997} For SF$_6$, the
experimentally measured polarizability of $4.49$\,\AA$^3$ was used and
evenly distributed over the fluorine atoms ($0.74$\,\AA$^3$ per F
atom).\cite{gussoni:1998}\\

The atomic van-der-Waals (vdW) parameters ($\epsilon_i,
R_{\mathrm{min},i}$) of N$_2$O were individually optimized by
least-squares fitting using 
the trust region reflective algorithm\cite{trf:1999} to best
describe the nonbonded interactions with Xe and SF$_6$ from comparing
with energies from CCSD/aug-cc-pVTZ calculations for a number of
N$_2$O--Xe and N$_2$O--SF$_6$ heterodimer structures. Lorenz-Berthelot
combination rules of atomic vdW parameters between atom types $i$
(solute) and $j$ (solvent) ($\epsilon_{ij} = \sqrt{\epsilon_i
  \epsilon_j}$ at an atom-atom distance of $R_{\mathrm{min},ij} =
R_{\mathrm{min},i}/2 + R_{\mathrm{min},j}/2$) were used
throughout. For the reference electronic structure calculations the
grid was defined by center of mass distances between N$_2$O and the
solvent with range $r=[2.0, 10.0]$\,\AA, and angles $\alpha=[0,
  180]^\circ$ with step size of $30^\circ$ between the N$_2$O bond
axis and the N$_2$O-solvent center of mass direction. Solute--solvent
dimer structures with interaction energies lower than 5\,kcal/mol
above the dimer minimum structure were considered which led to 102
conformations for N$_2$O--Xe whereas for N$_2$O--SF$_6$ there were 203
structures. Reference interaction energies for fitting the vdW
parameters were determined as follows. The total energy of the
N$_2$O--solvent (Xe or SF$_6$) pair with the largest separation was
considered to be the zero of energy and energies for all other dimer
structures were referred to this reference. Interaction energies for
the equilibrium pair structure were found with $-0.84$\,kcal/mol for
N$_2$O--Xe and $-1.46$\,kcal/mol for N$_2$O--SF$_6$. Correcting for
the basis set superposition error (BSSE) through counterpoise
correction (CPC)\cite{cpc} reduces the interaction energies by up to
$\sim 50\%$ to $-0.36$\,kcal/mol and $-0.85$\,kcal/mol,
respectively. The alternative ``chemical Hamiltonian
approach''\cite{mayer:1983} was found to yield similar
results\cite{paizs:1998} as the CPC which, however, is not
recommended for correlated wave function methods such as
CCSD.\cite{baerends:2014,herbert:2022} In addition, the corrections
due to BSSE are of a similar magnitude as the error in fitting the
van der Waals parameters. Therefore, it was decided to not correct
the interaction energies for BSSE. Then, the vdW parameters for
each atom $i$ of N$_2$O were optimized to match the interaction energy
predicted by CHARMM with the reference interaction energies from
electronic structure calculations for the respective dimer
conformations. The optimized vdW parameters are given in Table
S2.\\

\subsection{Molecular Dynamics Simulations}
Molecular dynamics simulations were performed with the CHARMM program
package.\cite{Charmm-Brooks-2009} Each system (N$_2$O in Xe and N$_2$O
in SF$_6$ at given temperature and solvent concentration) was
initially heated and equilibrated for $100$\,ps each, followed by
10\,ns production simulations in the $NVT$ ensemble using a time step
of 1\,fs for the leapfrog integration scheme. The N$_2$O/Xe systems
were simulated at a temperature of 291.2\,K and for N$_2$O/SF$_6$ the
temperature was 321.9\,K which both are slightly above the
experimental critical temperatures for condensation of xenon and
SF$_6$, respectively ($T_c(\mathrm{Xe}) = 289.74$\,K,
$T_c(\mathrm{SF_6}) = 318.76$\,K).\cite{ziegler:2019,ambrose:1987,haynes:2014crc}
A Langevin thermostat (coupling $0.1$\,ps$^{-1}$) was used to maintain
the temperature constant but was applied only to the solvent (Xe and
SF$_6$) atoms. Positions and velocities of snapshots of the
simulations were stored every $1$\,fs for analysis. As intermolecular
vibrational energy transfer is slow,\cite{ziegler:2022} the structure
of N$_2$O was optimized and new velocities from a Boltzmann
distribution at the simulation temperature were assigned to N$_2$O
after the heating step. This ensures that the kinetic energies along
the asymmetric, symmetric and bending modes match the thermal energy
with respect to the target simulation temperature. \\

The different simulation systems were prepared according to the
conditions used in the
experiments.\cite{ziegler:2018,ziegler:2019,ziegler:2022} Table
S1 summarizes the concentration $c$(N$_2$O) of N$_2$O,
molar volumes $V_m$ and critical density ratio $\rho^* =
\rho/\rho_c$. The experimentally determined critical densities are
$\rho_c = 1.11$\,g/ml for xenon and $\rho_c = 0.74$\,g/ml for SF$_6$
from which critical concentrations of $8.45$\,M and $5.06$\,M for
xenon and SF$_6$ are obtained,
respectively.\cite{ambrose:1987,haynes:2014crc} In all setups, the
simulation box contains one N$_2$O molecule and 600 Xe atoms or 343
SF$_6$ molecules which corresponds to similar simulation box volumes
for similar relative density ratios of the two solvents. In the
original parametrization study a simulation box containing 343 SF$_6$
molecules was used to fit temperature-pressure
properties.\cite{samios:2010} \\

In the MD simulations for N$_2$O in SF$_6$, electrostatic and
polarization interactions were only computed between the N$_2$O solute
and SF$_6$ solvent. Electrostatic and polarization contributions to
the SF$_6$ solvent-solvent interactions were neglected. Such a
procedure ensures that the pure (liquid, gas) properties of the
solvent are unaltered. All force field parameters are listed in Table
S2 in the supplementary information.  \\

\subsection{Analysis}
The line shape $I(\omega)$ of the IR spectra for N$_2$O in different
solvent densities are obtained via the Fourier transform of the
dipole-dipole correlation function from the dipole sequence of the
single N$_2$O molecule
\begin{equation}
  I(\omega) n(\omega) \propto Q(\omega) \cdot \mathrm{Im}\int_0^\infty
  dt\, e^{i\omega t} 
  \sum_{i=x,y,z} \left \langle \boldsymbol{\mu}_{i}(t)
  \cdot {\boldsymbol{\mu}_{i}}(0) \right \rangle
\label{eq:IR}
\end{equation}
A quantum correction factor $Q(\omega) = \tanh{(\beta \hbar \omega /
  2)}$ was applied to the results of the Fourier
transform.\cite{marx:2004} This procedure yields lineshapes but
not absolute intensities. For direct comparison, individual spectra
are thus multiplied with a suitable scaling factor to bring
intensities of all spectra on comparable scales.\\

The response of the solute on the solvent structure and dynamics was
evaluated by determining the frequencies of the quenched normal modes
(QNM) of N$_2$O for frames every 5\,fs of the simulation. For QNM a
steepest descent geometry optimization for N$_2$O within a frozen
solvent conformation for either 100 steps or until a gradient root
mean square of $\leq 10^{-4}$\,kcal/mol/\AA\/ was reached was carried
out. The $9 \times 9$ mass weighted Hessian matrix of N$_2$O
solute in the solvent with regard to the solute atom displacements
in Cartesian coordinates was determined and diagonalized to obtain
time series of all 9 normal modes including the effect of
solvent.\cite{stratt:1994,keyes:1996} Instantaneous normal mode
(INM) analyses were also applied on the solute by diagonalizing its
mass weighted Hessian matrix without prior geometry optimization on
N$_2$O to obtain insights on the impact of the solvent structure on
the translational and rotational modes of the N$_2$O
solute.\cite{stratt:1994}\\

The normalized vibrational frequency-frequency correlation
function (FFCF) $\left \langle \nu_\mathrm{as}(t) \cdot
\nu_\mathrm{as}(0) \right \rangle$ / $\left \langle \nu_\mathrm{as}(0)
\cdot \nu_\mathrm{as}(0) \right \rangle$ was computed for the time
series of the asymmetric stretch frequency $\nu_\mathrm{as}$ gathered
from the QNM analyses. The amplitude $A$, lifetimes $\tau_i$ and
offset $\Delta$ of a bi-exponential function $c(t) =
A\mathrm{e}^{-t/\tau_1} + (1-A)\mathrm{e}^{-t/\tau_2} + \Delta$ were
optimized to fit the normalized FFCF for $t \in [0.1,2.0]$ ps.\\

Radial distribution functions $g(r)$ for solvent-solvent and
solvent-solute pairs were determined from the average number of
molecules $\langle \mathrm{d}n_r \rangle$ of type B within the shell
in the range of $r - \Delta r /2 < r \le r + \Delta r /2 $ around
molecules of type A.
\begin{equation}
  g(r) = 
  \dfrac{1}{\langle \rho_\mathrm{local} \rangle} \cdot
  \dfrac{\langle \mathrm{d}n_r \rangle}{4\pi r^2 \Delta r}
\label{eq:rdf}
\end{equation}
Here, $\langle \rho_\mathrm{local} \rangle$ is the local density of
compound B around compound A within a range that is half the
simulation box edge length and the shell width was $\Delta r =
0.1$\,\AA. The average coordination number $\langle N(r') \rangle$ of
compounds B within range $r'$ around compound A can be obtained from
$g(r)$ according to
\begin{equation}
  \langle N(r') \rangle = 
  4\pi \langle \rho_\mathrm{local} \rangle
  \int_{0}^{r'} g(r) r^2 dr
\label{eq:ncr}
\end{equation}
\\

\section{Results}

\subsection{Validation of the Interaction Potentials}
First, the quality of the intramolecular PES for N$_2$O is discussed,
followed by a description of the van der Waals parameters for the
N$_2$O solute fit to the \textit{ab initio} reference calculations.\\

The RKHS model provides a full-dimensional, intramolecular PES for
N$_2$O which was originally developed for investigating the N+NO
collision reaction dynamics.\cite{koner:2020} The Pearson coefficient
$R^2$ of the RKHS representation and the full set of reference values
is $0.99983$ and the root mean squared error (RMSE) between RKHS and
reference energies up to 20\,kcal/mol above the equilibrium structure
(78 reference energies) is $0.13$\,kcal/mol.  To establish the
spectroscopic quality of the PES, the evaluation by the discrete
variable representation (DVR) method using the DVR3D\cite{dvr3d:2004}
package yields a fundamental asymmetric stretch frequency of N$_2$O of
$2229$\,cm$^{-1}$ compared with 2224\,cm$^{-1}$ from experiments in
the gas phase.\cite{herzberg:1945,herzberg:1950,kagann:1982} The
bending and symmetric stretch frequencies obtained from the DVR3D
calculations are 598\,cm$^{-1}$ and 1291\,cm$^{-1}$, respectively, and
the overtone of the bending mode lies at 1184
cm$^{-1}$. Experimentally, the bending and symmetric stretch are found
at 589\,cm$^{-1}$ and 1285\,cm$^{-1}$, respectively, while the
overtone of the bending frequency is at 1168
cm$^{-1}$.\cite{herzberg:1945,herzberg:1950,kagann:1982}\\

\begin{figure}
\includegraphics[width=0.70\textwidth]{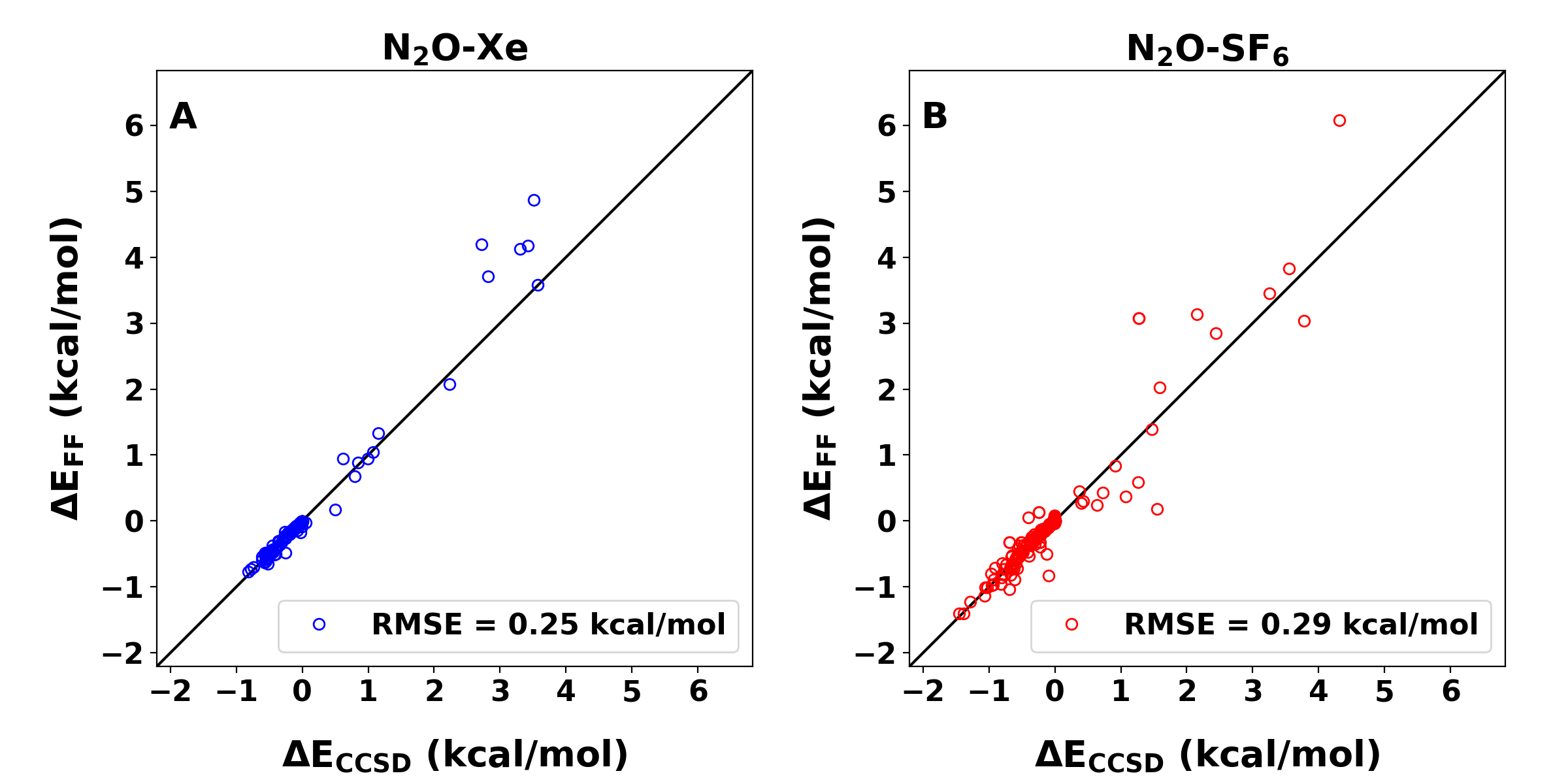}
\caption{\label{fig:1}
  Correlation between the CCSD/aug-cc-pVTZ reference
  interaction energies for the (A) N$_2$O--Xe pair for 102 different
  conformations and (B) N$_2$O--SF$_6$ pair for 203 different
  conformations with interaction energies lower than 5\,kcal/mol above
  the minimum, respectively.}
\end{figure}

Using MDCMs for the electrostatics of N$_2$O in the two solvent
environments (Xe and SF$_6$) requires a readjustment of the solute vdW
parameters. To keep the solvent-solvent interaction unchanged, only
the vdW parameters of N$_2$O were optimized with respect to
interaction energies from reference electronic structure
calculations.\\

Figure \ref{fig:1} reports the correlation between reference and
model interaction energies with optimized vdW parameters for N$_2$O
for the N$_2$O--Xe (panel A) and N$_2$O--SF$_6$ (panel B) complexes.
The reference interaction energies $\Delta E_\mathrm{CCSD} =
E_\mathrm{pair} - E_\mathrm{solute} - E_\mathrm{solvent}$ are the
differences between the respective total \textit{ab initio} energy of
the pair and the energies of the solute $E_\mathrm{solute}$ and
solvent $E_\mathrm{solvent}$ fragment in a fixed minimum energy
conformation.  The force field energies $\Delta E_\mathrm{FF} =
E_\mathrm{elec} + E_\mathrm{LJ}$ include electrostatic and
polarization contributions, and the Lennard-Jones potential between
the solute and solvent at the same conformation as in the reference
data set.  The energies sample attractive (negative energy) and
repulsive (positive energy) ranges of the N$_2$O--solvent energies and
the comparison is restricted to energies within 5\,kcal/mol of the
separated molecules which is the zero of energy. For N$_2$O--Xe, the
RMSE is $0.25$\,kcal/mol and for N$_2$O--SF$_6$ it is
$0.29$\,kcal/mol. The most stable structure is stabilized by
$-1.41$\,kcal/mol ($-1.37$\,kcal/mol from the fitted energy function)
for N$_2$O--SF$_6$, and $-0.80$\,kcal/mol ($-0.75$\,kcal/mol) for
N$_2$O--Xe.\\

The critical concentration at which transition to a supercritical
fluid occurs is another relevant property for the present
work. Earlier work showed that this transition can be correlated with
a pronounced increase in the local solvent reorganization lifetime
$\tau_\rho$.\cite{tucker:2000a,tucker:2000b} The quantity $\tau_\rho$
is the integral over the local-density autocorrelation function
$\tau_\rho = \int_{0}^{\infty} C_{\rho}(t) \mathrm{d}t$ and
characterizes the time required for the local environment around a
reference particle, e.g. the solute N$_2$O in the present case, to
change substantially. Hence, $\tau_\rho$ can also be considered a
local-density reorganization time.\cite{tucker:2000b}\\

Figures S1 and S2 report the local solvent
reorganization lifetimes $\tau_\rho$ for pure xenon and SF$_6$,
respectively, from MD simulations. The solvent environments are
defined by cutoff radii which correspond approximately to the first
and second minima of the solvent-solvent RDF, see Figure
S4. For SF$_6$ the force field was
parameterized\cite{samios:2010} to reproduce the experimentally
measured critical density with corresponding concentration
$c_\mathrm{crit}({\rm SF_6}) = 5.06$\,M at the critical
temperature.\cite{ambrose:1987,haynes:2014crc} The present simulations
yield a peak for $\tau_\rho$ at $c({\rm SF_6}) = 5.02$\,M, in close
agreement with the experimental critical concentration at the critical
temperature and the concentration with the local peak in $\tau_2$ in
Figure \ref{fig:4}D. For xenon, however, the parametrization of the
PES\cite{aziz:1986} did not include phase transition properties to the
supercritical regime. Figure S1 shows that the maximum in
$\tau_\rho$ occurs at $c({\rm Xe}) = 5.19$\,M. This compares with a
critical concentration of 8.45 M at $T_c$ from
experiment.\cite{ambrose:1987,haynes:2014crc}\\

\subsection{Infrared Spectroscopy}
\begin{figure}[htb!]
\includegraphics[width=0.80\textwidth]{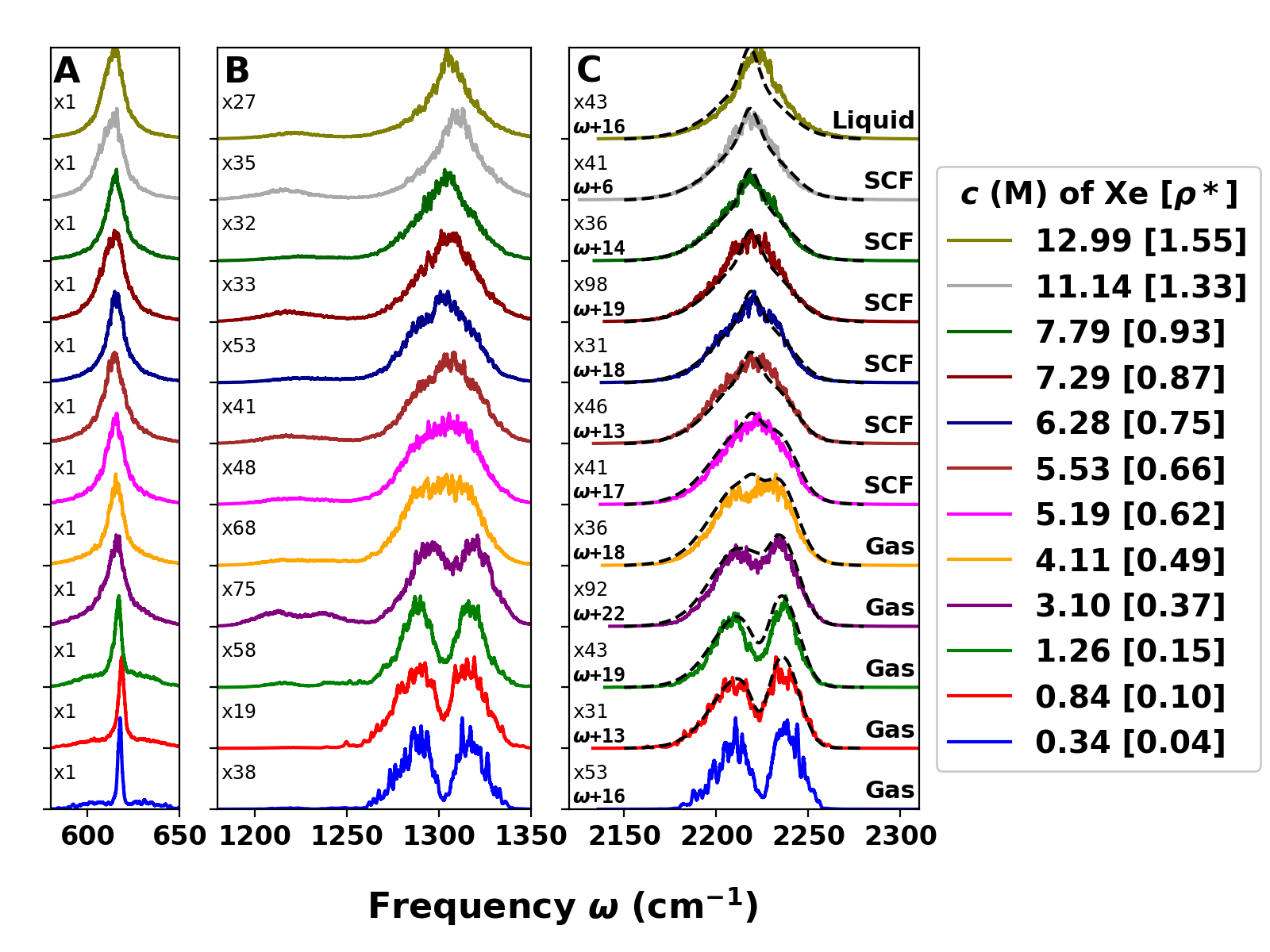}
\caption{\label{fig:2}  
  IR spectra of N$_2$O in xenon at different solvent
  concentrations for frequency ranges of (A) 580-650\,cm$^{-1}$, (B)
  1180-1350\,cm$^{-1}$ and (C) 2120-2310\,cm$^{-1}$ at $291.2$\,K.
  To make all spectra comparable, the amplitudes are scaled by
  the factor indicated at the center left for each range and
  density. In panel C the line shape frequency bands are shifted in
  frequency $\omega$ to maximize the overlap with the experimental IR
  signal (dashed black lines, at 291\,K for gas, SCF, and at 287\,K
  for liquid xenon) at the corresponding density for the N$_2$O
  asymmetric stretch vibration.  The frequency shift is also given in
  the bottom left corner, and an average shift is applied for
  densities without experimental reference spectra.}
\end{figure}

\begin{figure}[htb!]
\includegraphics[width=0.80\textwidth]{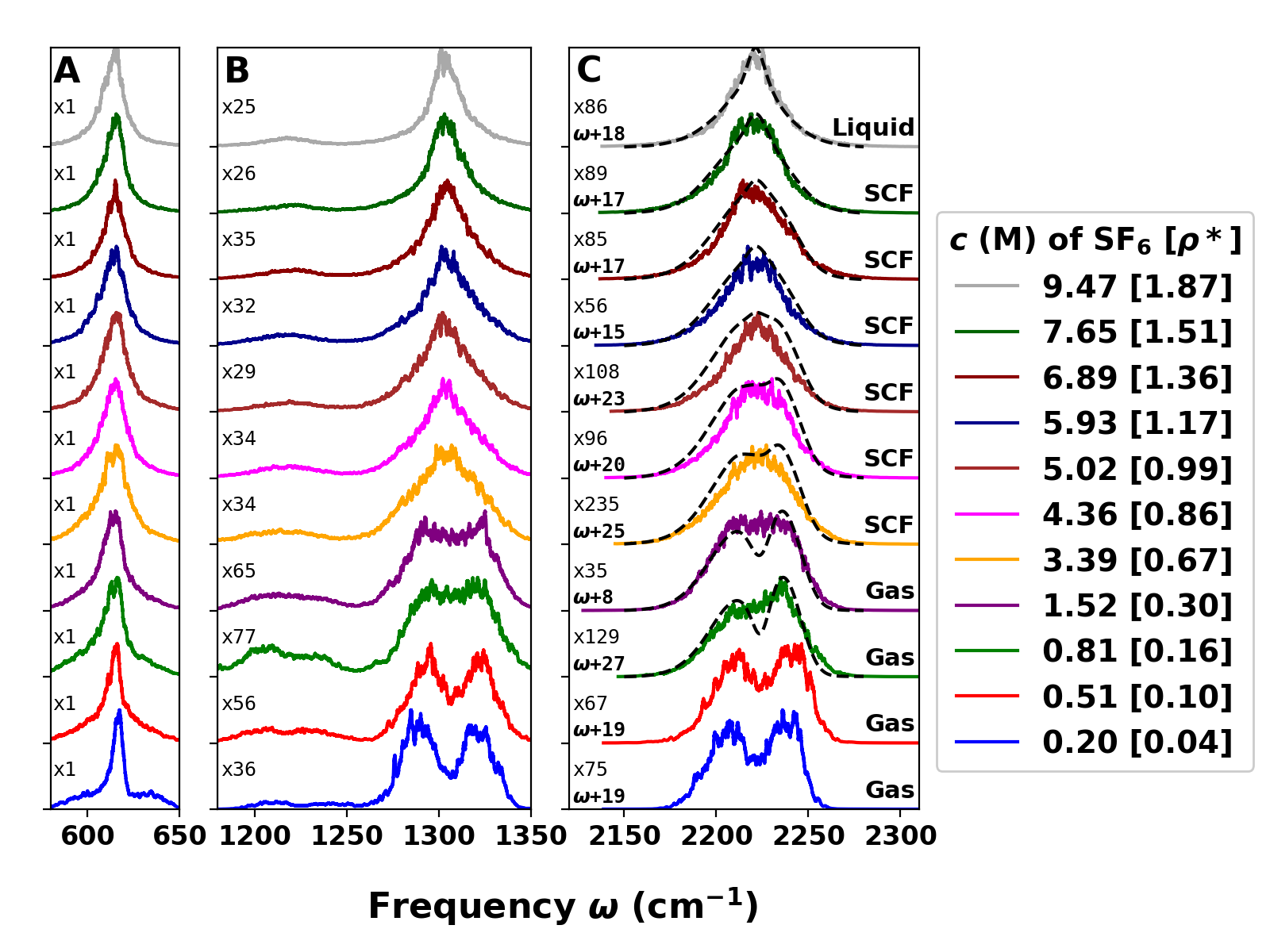}
\caption{\label{fig:3}
  IR spectra of N$_2$O in SF$_6$ at different solvent
  concentrations frequency ranges of (A) 580-650\,cm$^{-1}$, (B)
  1180-1350\,cm$^{-1}$ and (C) 2120-2310\,cm$^{-1}$ at $321.9$\,K.
  Further information are provided in the caption of Figure
  \ref{fig:2}. The experimental IR signal (at 322\,K for gas, SCF,
  and at 293\,K for liquid SF$_6$) at the corresponding density for
  the N$_2$O asymmetric stretch vibration are shown by the dashed
  black line.}
\end{figure}

The computed IR spectra for the N$_2$O asymmetric $(\nu_{\rm as})$
stretch in xenon and SF$_6$ solution as a function of solvent density
for near critical isotherms allows comparison between experimental and
simulation results.  The present work focuses mainly on the change of
the IR line shape at different solvent concentrations especially the
P-, Q-, and R-branch structure, see Figures \ref{fig:2} and
\ref{fig:3}. The P- and R-branches are the IR spectral features at
lower and higher wavenumber from the pure vibrational transition
frequency for the excitation of mode $\nu$. The $\nu_{\rm as}$,
$\nu_{\rm s}$, and $\nu_{\delta}$ band shapes arise due to
conservation of angular momentum during a vibrational excitation upon
photon absorption.\cite{herzberg:1945} The Q-branch is the absorption
band feature at the transition frequency. In addition to the
asymmetric stretch $\nu_\mathrm{as}$ ($\sim 2220$\,cm$^{-1}$, symmetry
A$_1$/$\mathrm{\Sigma^+}$), the symmetric stretch $\nu_\mathrm{s}$
($\sim 1305$\,cm$^{-1}$, A$_1$/$\mathrm{\Sigma^+}$) and the bending
vibration $\nu_\mathrm{\delta}$ ($\sim 615$\,cm$^{-1}$,
E$_1$/$\mathrm{\Pi}$) fundamentals from MD simulations are reported
and analyzed as a function of solvent density.
\\

From a quantum mechanical perspective, the rotational structure of an
IR-active vibrational mode for a linear molecule (such as N$_2$O) in
the gas phase leads to P- and R-branches. Selection rules dictate the
change of the rotational and vibrational quantum states, $j$ and
$\nu$, satisfying $\Delta \nu = \pm 1$ and $\Delta j = \pm 1$ for
vibrational modes of A$_1$/$\mathrm{\Sigma^+}$ symmetry, and $\Delta j
= 0,\pm 1$ for vibrational modes of E$_1$/$\mathrm{\Pi}$ symmetry,
respectively. For N$_2$O, the asymmetric $\nu_\mathrm{as}$ and
symmetric stretch $\nu_\mathrm{s}$ are vibrational modes with parallel
vibrational transition dipole moment to the bond axis and of
A$_1$/$\mathrm{\Sigma^+}$ representation, and the bending mode
$\nu_\mathrm{\delta}$ with perpendicular vibrational transition dipole
moment of E$_1$/$\mathrm{\Pi}$ representation. These rotational
selection rules are only strictly valid in the absence of
perturbations (exact free rotor) and break down with increasing
deviation from a free rotor model which, for example, can be due to
embedding the rotor into a solvent of different density which is a
means to tune the strength of the perturbation. As a consequence of
the perturbation a Q-branch-like spectral feature emerges for parallel
bands and becomes dominant with a band maximum at the wavenumber of
the corresponding pure vibrational transition energy at sufficiently
high solvent density.\\

For N$_2$O in xenon the IR band structure of $\nu_\mathrm{as}$ in
Figure \ref{fig:2}C ranges from resolved P- and R-branch in gaseous
xenon up to a single near-Lorentzian-shaped band in liquid xenon in
these classical MD simulations. From the simulations, at xenon
concentrations higher than $5.19$\,M, or relative density of $\rho^* =
\rho/\rho_c = 0.62$, the xenon solvent is a supercritical fluid. In
the simulated absorption spectra at this density and above the
$\nu_\mathrm{as}$ P- and R-band structure is not resolved and a
centrally peaked band at the pure vibrational frequency dominates the
band shape. The black dashed lines in Figure \ref{fig:2}C are the
experimentally determined spectra at the given solvent
concentration.\cite{ziegler:2022,ziegler:2019,ziegler:2018}\\

The agreement between experimental and computed N$_2$O in Xe
absorption band shapes is good and captures the dramatic gas
phase-to-condensed phase lineshape change as a function of solvent
density. However, the frequency position of the $\nu_\mathrm{as}$ band
needs to be blue-shifted by a small frequency shift $\omega$ to
achieve the best overlap with the experimental IR line shape.  The
density dependent shifts are indicated in the panel and range from
$+6$ to $+22$\,cm$^{-1}$ with no discernible trend. The shift
originates from different effects, including insufficient sampling of
the amount and distribution of internal energy within the N$_2$O
solutes vibrational degrees of freedom, remaining small inaccuracies
in the intermolecular interactions, neglecting many-body
contributions, and slightly underestimating the anharmonicity in the
PES along the relevant coordinates.\\

Figure \ref{fig:2}B shows the corresponding band shape for
$\nu_\mathrm{s}$ of N$_2$O in Xe around 1305\,cm$^{-1}$ in comparison
to experimentally measured 1291\,cm$^{-1}$ for N$_2$O in the gas
phase.  Similar to $\nu_\mathrm{as}$, it also displays resolved P- and
R-branches at low solvent concentrations and changes into a
Q-branch-like dominated structure in supercritical and liquid xenon.
The IR band of the N$_2$O bending mode $\nu_\mathrm{\delta}$ appears
at around $615$\,cm$^{-1}$ shown in Figure \ref{fig:2}A, compared
with 589\,cm$^{-1}$ from experiment in the gas
phase.\cite{herzberg:1945} At low solvent concentrations the band
structure of $\nu_\mathrm{\delta}$ from the MD simulations is a sharp
Q-branch with weaker P- and R-branch side bands consistent with the
quantum mechanical selection rules. The intensity of these bands are
no longer evident relative to the central bending feature at higher
solvent concentration. The first overtone of the $\nu_\mathrm{\delta}$
vibration is also detected between $1220$ to $1230$\,cm$^{-1}$ in
Figure \ref{fig:2}B but with low intensity. The computed bending
overtone $(2 \nu_{\rm b})$ exhibits the same P- and R-branches at low
density and the change to a single peaked band at high solvent
concentration as observed for the $\nu_\mathrm{s}$ and
$\nu_\mathrm{as}$ band structures, see Figures S5 and
S6, in agreement with experimental
results.\cite{nist}\\

The IR band structure of $\nu_\mathrm{as}$ of N$_2$O in SF$_6$ is
shown in Figure \ref{fig:3}C.  At the lowest solvent
concentrations $0.20$\,M and $0.51$\,M ($\rho^* = \{ 0.04, 0.10\}$)
the simulated distinct P- and R-branch structure is evident but
gradually disappears for $[\rm{SF_6}] > 1.52$\,M ($\rho^* > 0.30$) for
corresponding experimental state points. As the solvent concentration
of SF$_6$ increases the band shape changes into a single peaked band
in agreement with the experimental IR spectra for
$[\rm{SF_6}]=7.65$\,M and $9.47$\,M ($\rho^* = \{ 1.51, 1.87\}$).  The
computed IR spectra for $0.81 \leq [\rm{SF_6}] \leq 5.93$\,M ($0.16
\leq \rho^* \leq 1.17$) do not exhibit the double peak structure or
are narrower compared to the experimental absorption spectra. The
frequency position of the $\nu_\mathrm{as}$ band is also adjusted by a
small frequency blue shift $\omega$ within the range of $+8$ to
$+27$\,cm$^{-1}$ to maximize the overlap with the experimental IR line
shape.\\

The line shapes for the $\nu_\mathrm{\delta}$ and $\nu_\mathrm{s}$
vibrational modes are shown in Figures \ref{fig:3}A and B as a
function of SF$_6$ concentrations. The $\nu_\mathrm{s}$ mode changes
from P-/R-branches at low solvent concentrations into a single
featured structure in supercritical and liquid SF$_6$. The calculated
IR band of $\nu_\mathrm{\delta}$ around $615$\,cm$^{-1}$ exhibits a
sharp Q-branch with resolvable weak P- and R-branch satellites only at
low SF$_6$ concentration.  The first overtone of the perpendicularly
polarized $\nu_\mathrm{\delta}$ mode is again detected between $1220$
to $1230$\,cm$^{-1}$ with low intensity and a band shape as observed
for the parallel polarized fundamental $\nu_\mathrm{s}$ and
$\nu_\mathrm{as}$ band structures.\\

\subsection{Vibrational Frequency-Frequency Correlation Function}

\begin{figure}[htb!]
\includegraphics[width=0.90\textwidth]{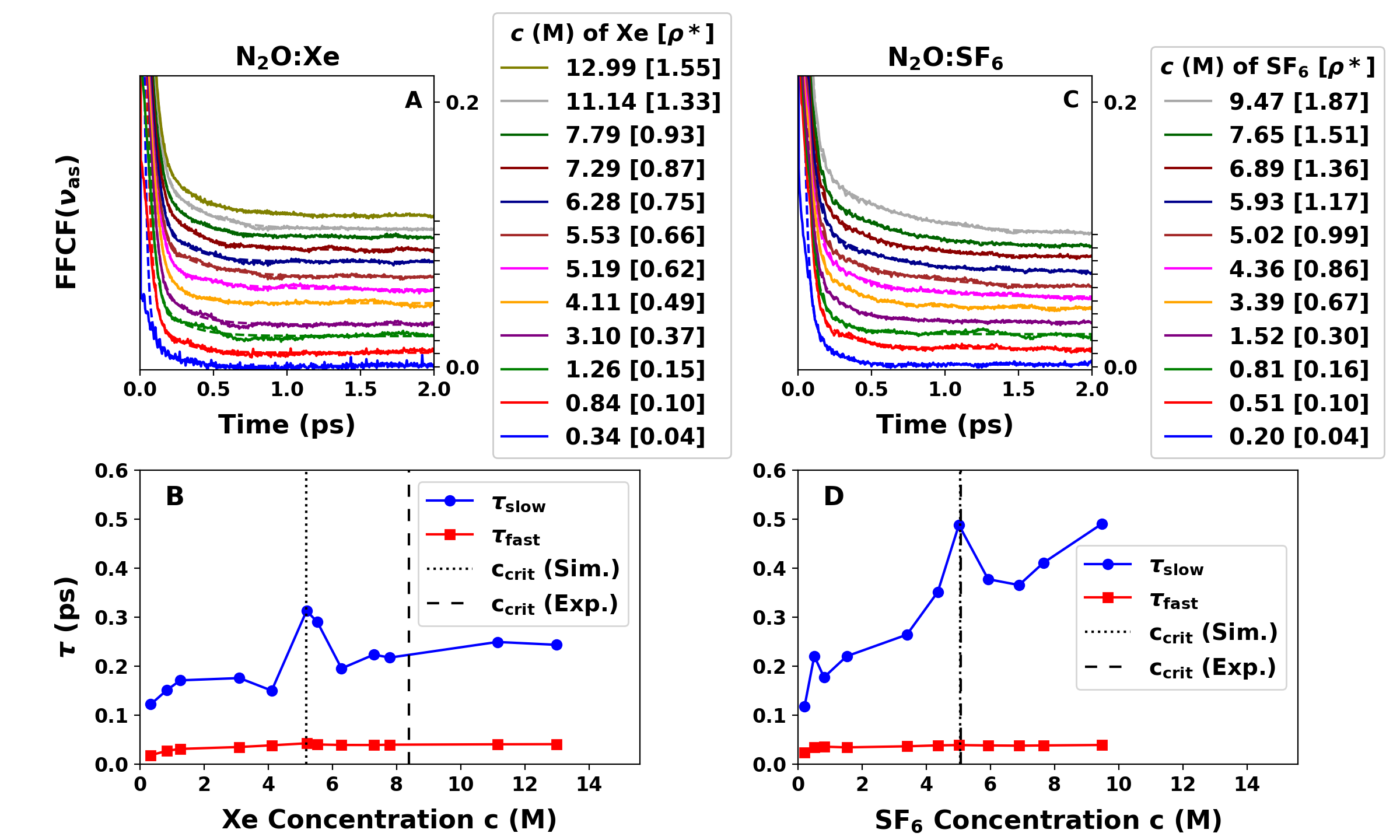}
\caption{\label{fig:4}
  FFCF of the asymmetric stretch frequencies $\nu_\mathrm{as}$
  of N$_2$O from QNM analysis in (A) xenon and (C) SF$_6$ solvent at
  different concentrations. Asymptotically, all FFCFs decay close to
  zero ($\Delta < 0.01$).  An offset of $0.01$ was added to the
  normalized FFCF to avoid overlap and the lifetimes $\tau$ from
  the fitted bi-exponential function are shown for the (B) xenon and
  (D) SF$_6$ solvent depending on solvent concentration. The critical
  concentration $c_\mathrm{crit}$ determined from experiments and its
  estimation from simulations at $T_c$ of xenon and SF$_6$ are marked
  by the vertical dashed and dotted lines, respectively.  It should be
  noted that the Xe--Xe interactions were not optimized to reproduce
  the experimentally known $T_c$ for pure Xe whereas the model for
  SF$_6$ does.}
\end{figure}

The vibrational frequency fluctuation correlation function (FFCF)
which probes the coupling of the solute vibrational modes to the
solvent environment can be determined from 2DIR experiments. Of
particular interest are the time scales of the lifetimes $\tau_i$ and
amplitudes $A_i$ with which the FFCF decays which are shown in Figure
S7.  In the condensed phase, the FFCF is also related to
the change of the center-line slope at different 2DIR waiting
times.\cite{hamm:2011} \\

Figures \ref{fig:4}A and C report the FFCFs for the asymmetric stretch
$\nu_\mathrm{as}$ mode of N$_2$O in xenon and SF$_6$ from
QNM.\cite{stratt:1996,MM.n3:2019} The fit of the FFCFs using a
bi-exponential decay yields two times: a rapid inertial component
($\tau_1 \sim 0.05$\,ps) and a longer spectral diffusion time scale
($\tau_2 \sim 0.2 \dots 0.5$\,ps). It is not possible to directly
compare the computed vibrational FFCFs from INM with the
experimentally measured 2DIR spectra as they are overwhelmingly
dominated by the contribution of rotational energy relaxation
dynamics.\cite{ziegler:2018,ziegler:2019,ziegler:2022} Nevertheless,
these time scales can be compared with measured and computed time
scales $[\tau_1,\tau_2,\tau_3]$ for CN$^-$ ($[0.2,2.9,{\rm n.a.}]$\,ps
and $[0.04,0.87,9.2]$\,ps) and for N$_3^-$ ($[{\rm n.a.}, {\rm
    n.a.},\sim 1.2]$\,ps and $[0.04,0.23,1.2]$\,ps) in
H$_2$O.\cite{hamm:2007,MM.cn:2013,maekawa:2005,MM.n3:2019} Here,
"n.a." refers to time scales that were not determined or available
from the experiments but clearly appeared in the analysis of the
simulations. A similar behaviour of the vibrational FFCF as the one
found in the present work - a first rapid relaxation time on the 0.05
to 0.1\,ps time scale, followed by a slower time scale of 0.5 to
1.0\,ps - was reported in an earlier MD simulation study for liquid
HCl.\cite{oxtoby:1983}\\

It is important to note that experimentally decay times on the several
10 fs time scale are difficult to determine with confidence from
vibrational FFCFs. Therefore only those longer than that are
considered in the following. It is noted that for both ions (CN$^-$
and N$_3^-$) in water, mentioned above, the longer decay times are 1\,ps 
or longer compared with $\sim 0.5$\,ps for N$_2$O in SF$_6$. This is
consistent with the fact that ion--water interactions are considerably
stronger than N$_2$O--SF$_6$ interactions. Furthermore, the N$_2$O--Xe
interaction is weakest among all those considered here which leads to
the rather short $\tau_2$ value for this system, even relative to
SF$_6$. Given that for the present system the intermolecular
interactions are weak and the decay times are rapid it is anticipated
that only $\tau_2$ for N$_2$O in SF$_6$ would be amenable to 2DIR
experiments.\\

Figures \ref{fig:4}B and D show the dependence of the short ($\tau_1$)
and long ($\tau_2$) time scales as a function of the xenon and SF$_6$
solvent concentration, respectively. In both solvents, the longer
lifetime slows down with increasing concentration and a peak in the
lifetime at $5.19$\,M in Xe and $5.02$\,M in SF$_6$ is observed. The
pronounced increase in $\tau_2$ is likely to be related to approaching
the critical point as the local peak in $\tau_2$ at a SF$_6$ solvent
concentration of 5.02\,M matches the critical concentration for pure
SF$_6$. Similarly, for N$_2$O in Xe the peak at $5.19$\,M is
consistent with a lengthening of the $\tau_2$ timescale in this
solution at the critical concentration for pure xenon which was
obtained from the local solvent reorganization lifetimes
$\tau_\rho$, see Figure S1. Thus, both solvents exhibit
critical slowing (lengthening of $\tau_2$) at the critical
concentration of the respective pure solvent.\\

It is also noted that the slope for $\tau_2(c)$ for N$_2$O in SF$_6$
is considerably steeper than for Xe as the solvent. This is most
likely also due to the increased solvent-solute interaction strength
between N$_2$O and SF$_6$ compared with Xe. The results for $\tau_2$
indicate that interesting dynamical effects, such as critical slowing,
can be expected to develop around the critical point of the
solvent. It is worthwhile to note that at the critical concentration
$c({\rm Xe}) = 5.19$\,M for the transition to a SCF the spectroscopy
of the asymmetric stretch changes from P-/R-branches to Q-branch-like
(Figure \ref{fig:2}) and the decay time $\tau_2$ from the FFCF of the
asymmetric stretch (see Figure \ref{fig:4}B) features a pronounced
maximum. The fact that for xenon the critical density from the
simulations is underestimated (5.19 M vs. 8.45 M from experiments) but
the concentration-dependence of the spectroscopy agrees with
experiment suggests that orientational relaxation and inhomogeneous
effects play the dominant role for the density dependence of the
dipole correlation function.\\

Direct comparison of the vibrational FFCFs from INM with the
prior experimental FFCFs is not meaningful at present because of the
overwhelming contribution to the reported FFCF is due to rotational
energy relaxation dynamics. 2DIR spectral features corresponding to
the vibrational FFCF overlap the rotational energy relaxation but
may be accessible in future experimental studies that employ higher
spectral resolution detection than in
Refs.\cite{ziegler:2018,ziegler:2019,ziegler:2022} Furthermore,
unlike typical 2DIR spectra of condensed phase systems, the
available experimental FFCFs report only on the quasi-free rotor
members of the ensemble contrary to the computed vibrational
FFCFs.\\

\subsection{Radial Distribution Functions and Size of the Solvent Shells}
A structural characterization of the solvent environment is afforded
by considering the radial distribution functions $g(r)$ between the
center of mass of N$_2$O and the central sulfur atom and the Xe atom,
which are shown in Figure S3A and C, respectively.
They show a wide first peak corresponding to the first solvation
shell around the N$_2$O solute within $\sim 4.3$\,\AA\/ (panel A) for
the xenon and $\sim 4.6$\,\AA\/ (panel C) for the SF$_6$ solvent,
followed by considerably weaker features around 8 \AA\/ and beyond
depending on the density.\\

\begin{figure}[htb]
\includegraphics[width=0.60\textwidth]{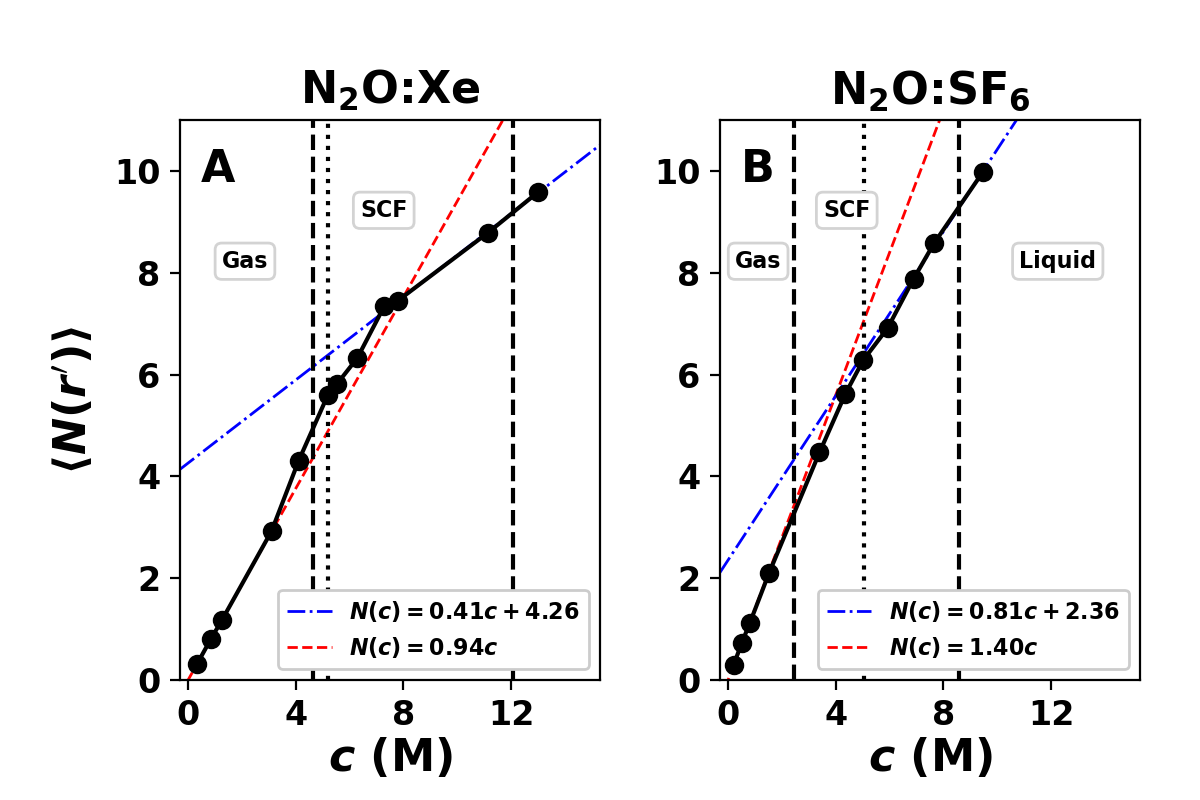}
\caption{\label{fig:5}
  Average coordination number $\langle N(r') \rangle$ of (A)
  the Xe ($r' = 6.5$\,\AA) and (B) SF$_6$ solvent compounds ($r' =
  7.5$\,\AA) within the first solvation shell around N$_2$O at cutoff
  distance $r'$. The red and blue lines are a linear fit of the first
  and last 3 data points, respectively.  The vertical dotted lines
  mark the estimation of the critical concentration in the simulations
  at $T_c$ of (A) xenon and (B) SF$_6$, respectively, and the vertical
  dashed lines mark the approximate concentrations of the solvent
  phase transition.}
\end{figure}

A more concise comparison of how the solvent environment depends on
solvent density can be obtained by considering the average
coordination number $\langle N(r') \rangle$ of the Xe and SF$_6$
solvent within the first solvation shell (Figures \ref{fig:5}A and B)
that are computed from the respective $g(r)$, see Figure
S3. At the respective lowest sampled solvent concentration,
$\langle N(r') \rangle$ yields an average of $0.31$ Xe atoms (panel A
at $0.34$\,M) and $0.29$ SF$_6$ molecules (panel B at $0.20$\,M)
within the first solvation shell.  The coordination number rises
monotonically with increasing solvent concentration but shows a
decrease in the slope within the SCF regime as the solvation shell
fills. It is interesting to note that both solvents exhibit two slopes
for $N(r)$ depending on $c$: a first, steeper one for the
gas-phase-like solvent into the SCF regime, and a second, flatter one,
leading into the liquid-like regime. The slopes for the gas-phase-like
regime is steeper (1.40\,M$^{-1}$) for SF$_6$ than for Xe
(0.94\,M$^{-1}$), which may be related to the solvent-solute
interaction strength.\\

It is also of potential interest to consider the present results
in the light of the independent binary collision model
(IBC)\cite{litovitz:1957,chesnoy:1984,adelman:1991} although the
model has also been criticized to be potentially oversimplified when
applied to liquids.\cite{oxtoby:1981,dardi:1991} In the simplest
implementation of the IBC model - where the contact distance is
associated with the position of the first maximum of the
solvent-solute radial distribution function which only
insignificantly changes with density - the rate for a given process,
e.g. vibrational relaxation, depends linearly on a) the collision
rate and b) a density independent probability that a given collision
is effective, see Figures S3A and C. As a consequence,
rates adequately described by an IBC model depend linearly on
density which itself is proportional to the occupancy of the first
solvation shell. This is what is found in the present work for low
densities, see Figure \ref{fig:5}. However, towards higher
densities, another linear dependence develops with lower slope for
both solvents. The breakdown of the IBC model occurs at slightly
higher solvent concentration for Xe than for SF$_6$ ($c({\rm Xe})
\sim 3.5$ M vs. $c({\rm SF_6}) \sim 2.9$ M, see Figure \ref{fig:5})
compared with values of $\sim 3.4$ M and 4.0 M from 2DIR
experimentally determined rotational energy relaxation,
respectively.\cite{ziegler:2022} Just as found for comparison of the
simulated and experimental absorption spectra, further refinement of
the N$_2$O--SF$_6$ interactions will provide even better
quantitative agreement with experiment, for example within a PES
morphing approach.\cite{MM.morph:99}\\

\section{Discussion}
Classical MD simulations have been previously used to determine and
analyze rotational vibrational spectra of simple solutes, including CO
or HCl in a solvent (Argon)\cite{berens:1981,bulanin:2010} or for
water.\cite{skinner:2019} These calculations demonstrated that
classical MD simulations are capable of realistically capturing P- and
R-branch structures in moderate to high density gases and even in
liquid water.\cite{skinner:2019} Along similar lines, the present MD
simulations of N$_2$O in xenon or SF$_6$ as the solvent reproduce well
the experimentally observed splitting in P-, Q- and R-branches and the
overall lineshapes in the IR spectra of N$_2$O depending on solvent
concentration (see Figures \ref{fig:2} and \ref{fig:3}). The
agreement in the computed line shapes of the FTIR $\nu_\mathrm{as}$
spectrum of N$_2$O in xenon does match with the experimental spectra
as a function of solvent concentration. The separation between the
maxima of the P- and R-branches is $\sim [27, 25, 29, 22]$\,cm$^{-1}$
at $[0.34, 0.84, 1.26, 3.10]$\,M from the simulations, compared with
$\sim [25, 25, 22]$\,cm$^{-1}$ at $[0.84, 1.26, 3.10]$\,M from the
experiments. The IR signals for the $\nu_\mathrm{as}$ vibration of
N$_2$O in SF$_6$ show a larger contribution of the Q-branch-like
feature at lower $\rho^*$ compared to that observed in the
experimental spectra. For simulations in both solvents the computed
spectra are shifted to the red compared with the results from
experiments for $\nu_\mathrm{as}$.\\

The computed IR spectra from the classical MD simulations of N$_2$O/Xe
in Figure \ref{fig:2} agree with the relative intensity and band
width of the $\nu_\mathrm{as}$ vibration signal at the corresponding
solvent concentration. The IR spectra of N$_2$O in SF$_6$ in Figure
\ref{fig:3} deviate more from the corresponding experimental line
shapes by overestimating solvent-solute interactions, leading to a
somewhat stronger contribution of the Q-branch-like absorption
character at lower solvent concentration. The experimental
$\nu_\mathrm{as}$ signal at the lowest measured solvent density of
$0.81$\,M overlaps better with the computed signal at lower solvent
densities of $0.51$\,M and $0.20$\,M. However, the line shape in
highly dense supercritical and liquid SF$_6$ agrees well with
experiment where the Q-band-like feature dominates the spectral
lineshape and resolvable P- and R-branch contributions are not
evident. At the same absolute density the asymmetric stretch of N$_2$O
is indicative of a more liquid-like character of SF$_6$ as the solvent
compared with xenon, see Figures \ref{fig:2}C and
\ref{fig:3}C.\\

Splitting of the IR spectrum into P- and R-branches correlates with
free rotor properties of the N$_2$O solute molecule. Based on this,
the results in Figure \ref{fig:3} indicate an overestimation of
the solute-solvent interaction in SF$_6$ because a merged line shape
for $\nu_\mathrm{as}$ arises between 0.81\,M and 1.52\,M from the
computations whereas experimentally, this occurs for concentrations
between 3.39\,M and 4.36\,M. Consequently, the rotation of N$_2$O in the
simulations starts to become hindered at lower solvent densities
compared to experiment. This observation correlates with the
long-range and stronger electrostatic interaction between N$_2$O and
polar SF$_6$ rather than xenon as the solvent.\\

As a comparison, the electrostatic contribution to the interaction
energy of a N$_2$O/SF$_6$ pair in its equilibrium conformation with
$-1.42$\,kcal/mol is considerably stronger then the polarization
contribution in an N$_2$O/Xe pair with $-0.84$\,kcal/mol. As the
electrostatic MDCM and vdW parameter fit are only applied to single
molecules or molecule pairs, respectively, the energy function might
insufficiently capture many-body interactions between the N$_2$O
solute and multiple SF$_6$ solvent molecules.\\

An overtone of the $\nu_\mathrm{\delta}$ vibration (615\,cm$^{-1}$) of
N$_2$O is found at about twice the frequency between 1220 to
1230\,cm$^{-1}$ in both solvents. The splitting into P- and R-branches
at low solvent densities originates from excitation of simultaneously
two bending mode quanta ($2\nu_\mathrm{\delta}$) and the change of the
rotational state by $\Delta J = \pm 1$.\cite{abouaf:2000,chuanxi:2021}
Consequently, as the splitting of the $2\nu_\mathrm{\delta}$ IR band
depends on the free rotor properties such as for the $\nu_\mathrm{as}$
and $\nu_\mathrm{s}$ IR band structure, the band shape changes towards
a single Q-branch peak at higher solvent concentrations. In other
words, the rotational ``selection rules'' followed by the bending
overtone are the same as for the asymmetric stretch vibration which
both have the same symmetry. This quantum mechanical result is seen in
the present classical MD simulations.\\

The perturbation of the rotational modes by the solvent are visualized
by the INM in Figures S8A and B for N$_2$O in SF$_6$ and
xenon at different densities, respectively.  The INM histogram for
N$_2$O in SF$_6$ shows two peaks around the zero frequency line for
low positive and imaginary frequencies (positive and negative second
derivative of the potential projected along normal mode displacement,
respectively). The emerging shoulder(s) in the INM histogram
towards higher solvent densities (Figure S8) are due to
low-frequency modes involving solute-solvent interactions. This
leads to larger deviations from a free rotor movement of the N$_2$O,
akin to motion in a density-dependent rotationally constraining
solvent potential.\cite{ziegler:2019} Visual inspection of the
normal modes identifies these as librational modes or frustrated
rotations. The INM histogram for N$_2$O in xenon in Figure
S8B shows less narrow peaks and lower shoulders for gaseous
solvent in comparison with the INM distribution of N$_2$O in SF$_6$ in
Figure S8A. The higher mode density around low frequencies
for the INMs of N$_2$O in xenon relative to SF$_6$ also indicates
lower perturbation of free rotor character in gaseous and
supercritical xenon solvent than SF$_6$.\\

The change of the IR spectra for linear molecules in gaseous,
supercritical up to liquid solvents can be compared with the
experimental and theoretical work on hydrogen cyanide in an electric
field.\cite{herschbach:1992,miller:1992} With increasing field
strength, the molecules evolve from a free rotor to those trapped in
pendular states which is one way to control the population of
rotational states in molecular beam experiments. The electric field
lifts the degeneracy of the quantum number $m_j$ (the projection of
the angular momentum on the electric field direction) for each
rotational state $j$ and leads to a manifold of different transitions
that corresponds to the IR spectra. Pendular states are linear
combinations of field-free $(j, m_j)$ states covering a range of
$j-$values but sharing the same $m_j$ quantum number, i.e. $j-$mixing
occurs.\cite{herschbach:1992} With increasing electric field strength
the field-free selection rule $\Delta j = \pm 1$ breaks down.  For
excitation along dipole transition vectors parallel to the electric
field the P- and R-branch change into a single Lorentzian-shaped
Q-branch at large electric field strength.\cite{herschbach:1992} \\

The effects found for polar molecules in external electrical fields
can be compared with the motion of N$_2$O in a supercritical or liquid
solvent. Here, the solvent molecules form a cavity around the solute
where the direction of the angular momentum is not energetically
degenerate and N$_2$O behaves like in a pendular state. In analogy to
a molecule in an electric field, the P- and R-branches collapse into a
single Q-branch-like band structure for vibrational modes with
transition dipole moments perpendicular to the angular momentum of the
molecule upon breakdown of the usual selection rules.\\

\begin{figure}[htb!]
\includegraphics[width=0.60\textwidth]{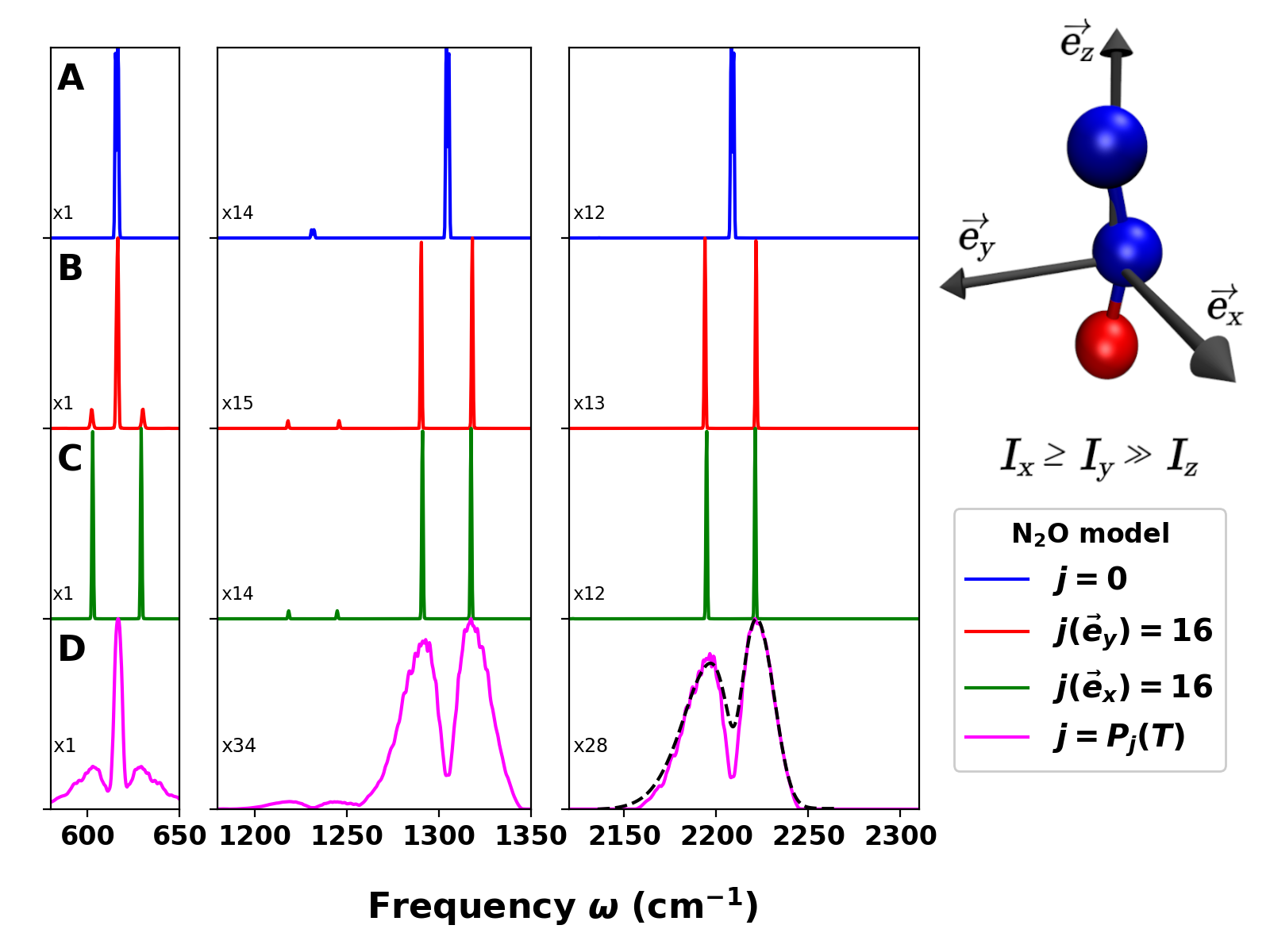}
\caption{\label{fig:6} 
  Model IR spectra from 100\,ps MD simulations of a single
  N$_2$O molecule with vibrational energy equivalent to $T=321.9$\,K
  and either (A) no rotational energy, rotation around axis (B)
  perpendicular ($\vec{e}_x$) or (C) parallel ($\vec{e}_y$) to the
  transition dipole moment of the bending vibration of N$_2$O. Panel D
  shows a superposition of N$_2$O IR spectra from MD simulation at
  different rotational states $j$ scaled by the respective probability
  according to the Maxwell-Boltzmann distribution $P_j$ at
  $T=321.9$\,K.  The experimental spectra of N$_2$O in gaseous SF$_6$
  ($c=1.26$\,M) is shown by the dashed line and shifted by
  $-14.5$\,cm$^{-1}$ to maximize the overlap with the computed line
  shape.  An illustration of an angles N$_2$O molecule with the
  principle axis of inertia are shown in the top right corner and the
  order of the moments of inertia $I$.}
\end{figure}

The IR spectra derived from classical MD simulations can also be
analyzed following a model by Skinner {\it et al.} who considered the
dipole-dipole correlation function $\phi = \phi_\mathrm{vib} \cdot
\phi_\mathrm{rot}$ as a product ansatz of a vibrational
$\phi_\mathrm{vib}$ and a rotational component
$\phi_\mathrm{rot}$.\cite{skinner:2019} The composition of the IR band
by P/R-branch and Q-branch depends on the angle $\theta$ between the
vibrational transition vector and the rotational axis. Assigning the
vibration and rotation with an angular frequency $\omega_\mathrm{vib}$
and $\omega_\mathrm{rot}$, the time-dependence of the correlation
function $\phi(t)$ for an idealized classical rotor (without damping)
has been expressed as
\begin{equation}
    \phi(t) = \cos \theta \cdot \exp^{(i \omega_\mathrm{vib} t)}
              + (1 - \cos \theta)  \cdot \exp^{(i (\omega_\mathrm{vib} \pm \omega_\mathrm{rot}) t)}
\label{eq:modelcorr}
\end{equation}
which assumes constant angular velocity, an approximation that does
not necessarily hold for vibrating molecules.\cite{skinner:2019} The
Fourier transform of $\phi(t)$ in equation \ref{eq:modelcorr} leads to
one Q-branch signal at $\omega_\mathrm{vib}$ for parallel alignment of
the rotational axis and vibrational transition dipole vector ($\cos
\theta=1$) and two P- and R-branch signals at $\omega_\mathrm{vib} -
\omega_\mathrm{rot}$ and $\omega_\mathrm{vib} + \omega_\mathrm{rot}$
for perpendicularly aligned rotational axis and vibrational transition
dipole vector ($\cos \theta=0$). For other cases or an unstable
rotation axis, signals for all P-, Q- and R-branches arise in the IR
spectra.  \\

Figure \ref{fig:6} shows computed IR spectra from the Fourier
transform of the dipole-dipole correlation function from MD
simulations of a single N$_2$O molecule with different initial
conditions.  For all cases (A-D) intramolecular energy is distributed
along all vibrational normal modes with respect to the thermal energy
at $321.9$\,K.  Without any rotational energy assigned the IR spectra
in Figure \ref{fig:6}A shows no split into P- and R-branches.
With rotational energy assigned around rotational axis {\it parallel}
to the transition dipole vector the IR spectra in Figure
\ref{fig:6}B shows separation into P- and R-branches by
28\,cm$^{-1}$ for $\nu_\mathrm{as}$ stretch that correlates for a
rotational constant of $B_{j=16} = 0.43$\,cm$^{-1}$. The IR signal of
the bending mode primarily consists of the Q-branch with two small P-
and R-branch satellites separated by $\pm 14$\,cm$^{-1}$ each from the
Q-branch. For rotational energy assigned around a rotational axis {\it
  perpendicular} to both stretch and bending transition dipole moments
all IR signals in Figure \ref{fig:6}C split in P- and R-branches
as well ($B_{j=16} = 0.42$\,cm$^{-1}$) but no Q-branch is
observed. The rotational energy corresponding to the rotational
quantum state $j=16$ is the state with the highest probability
according to the Maxwell-Boltzmann distribution function $P_j(T)$ for
N$_2$O at $321.9$\,K.\\

At lower solvent density, the appearance of P-, Q- and R-branch in the
model IR spectra in Figure \ref{fig:6}B can be explained by the
mechanical instability of a rotation around the transition dipole
vector of the bending mode $e_y$. For bent N$_2$O the moment of
inertia are ordered by $I_y > I_y \gg I_z$. According to the tennis
racket theorem\cite{tennis_racket} the rotation around the axis $e_y$
with the middle moment of inertia is unstable. Thus, the angle between
rotation axis or angular momentum vector and the transition dipole
vector of the bending mode varies in time and gives rise to P-, Q- and
R-branch signals.  \\

A superposition of model IR spectra at varying rotational energy
states and randomly sampled rotation axis but perpendicular to the
N$_2$O bond axis yields a line shape close to the computed spectra of
N$_2$O in gaseous solvent at lower concentrations. The contribution of
the single IR spectra at rotational state $j$ to the model spectra in
Figure \ref{fig:6}D are scaled by the probability value
according to $P_j(T=321.9$\,K$)$. An average rotational constant $B =
0.42$\,cm$^{-1}$ is computed from the spectra with rotational states
$j=[2,26]$, which is in perfect agreement of the experimentally
determined rotational constant of $B_\mathrm{exp} =
0.419$\,cm$^{-1}$.\cite{Tidwell:60} Hence, overall two regimes can be
distinguished: for low densities the rotational motion of the solute
is oscillatory and guided by the mechanical instability around the
middle rotational axis (tennis-racket theorem) whereas at higher
densities the constraining effect of the solvent due to tighter
packing leads to a more diffusive rotational motion of the solute.\\

\begin{figure}[htb!]
\begin{center}
  \includegraphics[width=0.90\textwidth]{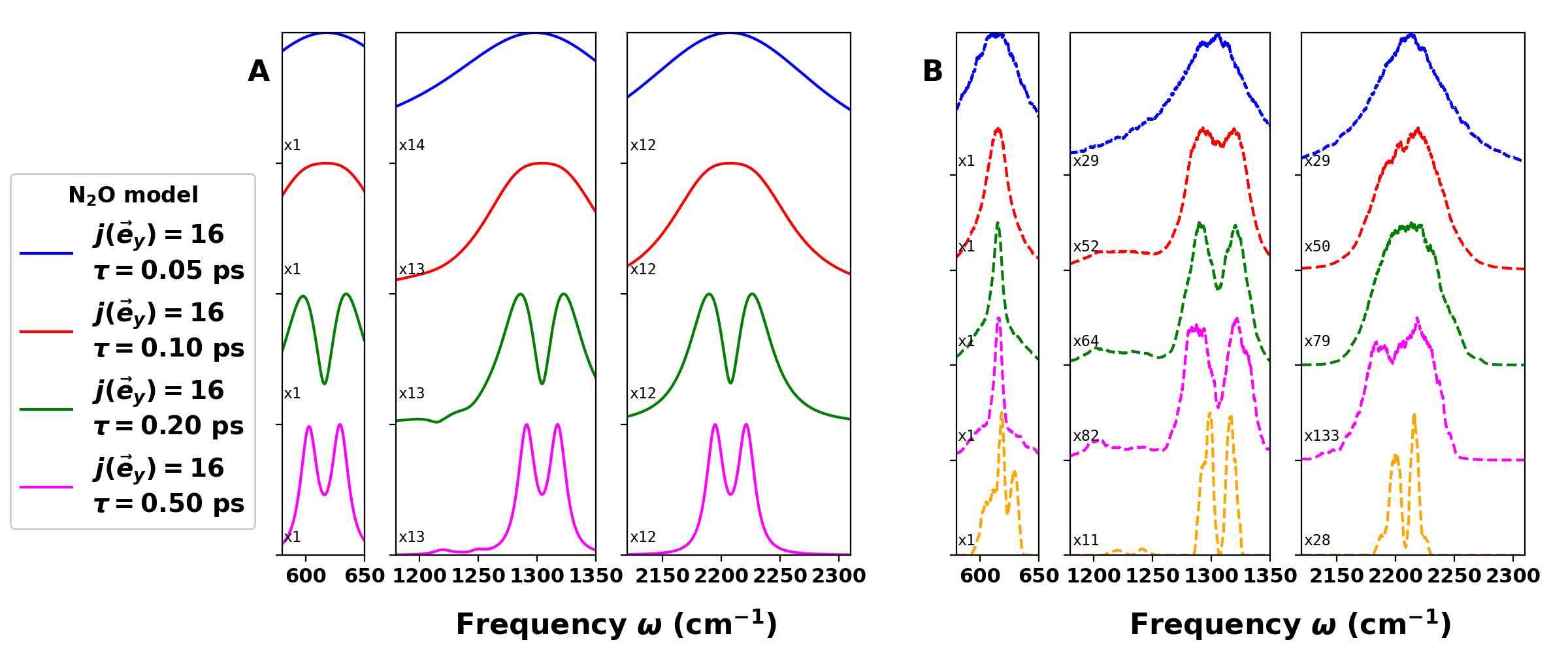}
\caption{Panels A: Model IR spectra from a Fourier transform of the
  dipole-dipole correlation function multiplied by a exponential
  damping factor with $\tau$ which describes orientational
  (rotational) relaxation and inhomogeneous effects. The dipole-dipole
  correlation function is obtained from a 100\,ps MD simulation of a
  single N$_2$O molecule with vibrational energy equivalent to
  $T=321.9$\,K and rotation around the axis parallel ($\vec{e}_y$) to
  the transition dipole moment of the bending vibration of
  N$_2$O. Panel B: Model IR spectra from 10\,ns Langevin simulations
  of a single N$_2$O molecule with initial vibrational and rotational
  energy equivalent to $T=321.9$\,K. A Langevin thermostat with
  different friction coefficients $\gamma_i = \{10, 1, 0.1, 0.01,
  10^{-4} \}$\,ps$^{-1}$ (from top to bottom) was applied to the
  dynamics to model energy exchange between N$_2$O and the solvent.}
\label{fig:7}
\end{center}
\end{figure}

For additional clarification for how the transition between
P-/R-features and Q-branch-like lineshapes occurs two different models
are briefly considered. The first one operates directly on the
dipole-dipole correlation function of a single N$_2$O molecule whereas
the second one uses a Langevin-type simulation of one N$_2$O molecule
in the gas phase. The influence of a solvent can be heuristically
included by multiplication of the dipole-dipole correlation function
$\phi(t)$, see Eq. \ref{eq:modelcorr}, with an exponential damping,
i.e. considering $\phi(t) \times \exp(-t/\tau)$. The damping with
characteristic time $\tau$ models all sources of the dipole
correlation function decay, primarily the solvent density dependence
on orientational relaxation due to collisions between N$_2$O and the
solvent in the present case. With increasing density of the solvent,
for example, $\tau$ decreases, which leads to loss of knowledge about
its rotational motion as well as broadening of the computed IR
signals, see Figure \ref{fig:7}A. Consequently, the
P-/R-features collapse into a single Q-branch-like peak for relaxation
decays $0.1 \leq \tau \leq 0.2$\,ps which are considerably shorter
than the rotational periods.\\

Alternatively, interaction with a solvent can be modelled from MD
simulations with a stochastic thermostat, such as in Langevin
dynamics, which applies random forces on the atoms to simulate
orientational interactions with the heat bath. For this, 10 ns
Langevin simulations with varying coupling strengths $\gamma$ were run
for a single N$_2$O molecule in the gas phase using CHARMM and the
dipole-dipole correlation function was determined. Figure
\ref{fig:7}B shows the computed IR spectra. At low coupling
strength $\gamma_i=10^{-4}$\,ps$^{-1}$ the IR spectra
still yield a split of the asymmetric stretch into P- and
R-branches (yellow trace). With increasing friction $\gamma_i=0.01$\,ps$^{-1}$
still multiple peaks are visible in the asymmetric stretch absorption
lineshape but the spectral Q-branch-like region is getting filled in
(pink trace). At yet higher $\gamma_i \ge 0.1$\,ps$^{-1}$
(green, red, blue traces) the P-/R-features wash out entirely.
\\

Finally, it is of some interest to discuss the results on the
spectroscopy (Figures \ref{fig:2} and \ref{fig:3}) and the solvent
structure (Figure \ref{fig:5}) from a broader perspective. The
interaction between N$_2$O and Xe is weaker by about 50 \% than that
between N$_2$O and SF$_6$. The type of interaction also differs: pure
van der Waals (in the present treatment) for Xe versus van der Waals
and electrostatics for SF$_6$. It is interesting to note that the
transition between the P-/R- and the Q-band-like spectra in Xe occurs
at $\rho^* \sim 0.6$ ($\sim 5.2$\,M) compared with $\rho^* \sim 0.2$
($\sim 0.9$\,M) for SF$_6$, see Figures \ref{fig:2} and
\ref{fig:3}. This is consistent with the increased interaction
strength for N$_2$O--SF$_6$ as fewer solvent molecules are required to
sufficiently perturb the system to illicit the change in the
spectroscopy. Similarly, the slope of $N(r)$ vs. concentration can
also serve as a qualitative measure for the interaction strength. To
stabilize a solvent shell of a certain size a smaller number (lower
concentration) of solvent molecules is required if the solvent:solute
interaction is sufficiently strong. These qualitative relationships
are also reflected in the longer $\tau_2$ values of the FFCFs for
N$_2$O in SF$_6$ compared with Xe as the solvent, see Figure
\ref{fig:4}. For large polyatomic molecules in highly
compressible fluids, ``attractive'' solutes have been found to
recruit increased numbers of solvent molecules which leads to local
density augmentation.\cite{song:2000} While the amount of
augmentation was found to correlate well with the free energy of
solvation of the solute from simulations, computations and
experiments were reported to follow different correlations.\\

\section{Conclusion}
In conclusion, the present work establishes that the rotational
structure of the asymmetric stretch band is quantitatively described
for N$_2$O in xenon and is qualitatively captured in SF$_6$ compared
with experiment. Specifically, the change from P-/R-branches in the
gas phase environment changes to a Q-branch-like structure, eventually
assuming a Lorentzian-like lineshape once the solvent passes through
the supercritical fluid to the liquid state. Transition between the
gas- and SCF-like solvent is reflected in a steep increase in the
longer correlation time $\tau_2$ in the FFCF for both solvents.
As the density of the solvent increases, additional solvent shells
appear in $g(r)$. The present work suggests that atomistic
simulations together with machine learned and accurate electrostatic
interactions yield a quantitative understanding of the spectroscopy
and structural dynamics\cite{MM.rev:2022} from very low density to the
liquid environment with the possibility to characterize transition to
the SCF.

\section{Supplementary Material}
See supplementary material for additional data and figures referred to
in the manuscript that further support the findings of this work.

\section{Acknowledgment}
This work was supported by the Swiss National Science Foundation
grants 200021-117810, 200020-188724, the NCCR MUST, and the University
of Basel (to MM) and by European Union's Horizon 2020 research and
innovation program under the Marie Sk{\l}odowska-Curie grant agreement
No 801459 -FP-RESOMUS which is gratefully acknowledged. The support of
the National Science Foundation grant CHE-2102427 (LDZ) and the Boston
University Photonics Center is gratefully acknowledged. We thank James
L. Skinner for valuable correspondence on Ref.\citenum{skinner:2019}.

\section{Data Availability}
Raw data were generated at the University of Basel large scale
facility.  Derived data supporting the findings of this study are
available from the corresponding author upon reasonable request. The
data and source code that allow to reproduce the findings of
this study are openly available at
\url{https://github.com/MMunibas/N2O_SF6_Xe}.

\bibliography{ref_rotvib}

\end{document}


\section{System Setup}

\begin{table}
\caption{System setup of N$_2$O concentration $c(\mathrm{N_2O})$,
  molar volumes $V_m$ and critical density ratio $\rho^*$ of one
  N$_2$O in 343 SF$_6$ molecules and one N$_2$O in 600 Xe atoms.}
  \label{sitab:conc}
\begin{tabular}{ccc||ccc}
\hline\hline
\multicolumn{3}{c||}{N$_2$O/SF$_6$} & \multicolumn{3}{c}{N$_2$O/Xe} \\
\hline
$c(\mathrm{N_2O})$ (mol/l) & $V_m(\mathrm{SF_6})$ (cm$^3$/mol) & $\rho^*$ &
$c(\mathrm{N_2O})$ (mol/l) & $V_m(\mathrm{Xe})$ (cm$^3$/mol) & $\rho^*$ \\
\hline
0.20 & 4934 & 0.04  & 0.34  & 2984 & 0.04  \\
0.51 & 1974 & 0.10  & 0.84  & 1194 & 0.10  \\
0.81 & 1234 & 0.16  & 1.26  & 796  & 0.15  \\
1.52 & 658  & 0.30  & 3.10  & 323  & 0.37  \\
3.39 & 295  & 0.67  & 4.11  & 244  & 0.49  \\
4.36 & 229  & 0.86  & 5.19  & 193  & 0.62  \\
5.02 & 199  & 0.99  & 5.53  & 181  & 0.66  \\
5.93 & 169  & 1.17  & 6.28  & 159  & 0.75  \\
6.89 & 145  & 1.36  & 7.29  & 137  & 0.87  \\
7.65 & 131  & 1.51  & 7.79  & 128  & 0.93  \\
     &      &       & 11.14 & 90   & 1.33  \\
     &      &       & 12.99 & 77   & 1.55  \\
\hline \hline
\end{tabular}
\end{table}

\begin{table}
\caption{Bonded and non-bonded parameters. Larger sets of 
parameters are made available on a github as full parameter files.}
\label{sitab:tab2}
\begin{tabular}{c|cc}
\hline\hline
\textbf{Residues} & \multicolumn{2}{c}{Parameters} \\
\hline \hline 
\textbf{N$_2$O} & \multicolumn{2}{c}{} \\
\hline
Intramolecular Potential & \\
RKHS\cite{koner:2020} & \multicolumn{2}{c}{\url{https://github.com/MMunibas/N2O_SF6_Xe/}} \\
 & \multicolumn{2}{c}{in, e.g., \url{N2O_Xe/source/pes1_rRz.csv}} \\
 & \multicolumn{2}{c}{or \url{N2O_SF6/source/pes1_rRz.csv}} \\
\hline 
Electrostatic Potential & \\
polarizable MDCM\cite{devereux:2020mdcm} & \multicolumn{2}{c}{\url{https://github.com/MMunibas/N2O_SF6_Xe/}} \\
 & \multicolumn{2}{c}{in, e.g., \url{N2O_Xe/source/n2o_xe.dcm}} \\
 & \multicolumn{2}{c}{or \url{N2O_SF6/source/n2o_sf6.dcm}} \\
\hline 
Polarizabilities &  $\alpha$ (1/\AA$^3$) &  \\
N1 & $0.94$ &  \\
N2 & $0.94$ &  \\
O & $0.94$ &  \\
\hline 
Non-bonded in SF$_6$ & $\epsilon_i$ (kcal/mol)& $R_{\mathrm{min},i}/2$ (\AA) \\
N1 & $0.2592$ & $1.716$ \\
N2 & $0.1542$ & $1.618$ \\
O  & $0.2089$ & $1.555$ \\
\hline 
Non-bonded in Xe & $\epsilon_i$ (kcal/mol)& $R_{\mathrm{min},i}/2$ (\AA) \\
N1 & $0.3775$ & $1.637$ \\
N2 & $0.0002$ & $2.565$ \\
O  & $0.3684$ & $1.497$ \\
\hline \hline
\textbf{SF$_6$} & \multicolumn{2}{c}{Samios and coworker\cite{samios:2010}} \\
\hline 
Bonds & $k_b$ (kcal/mol/\AA$^2$) & $r_\mathrm{min}$ (\AA) \\
S-F & $165.746$ & $1.565$ \\
\hline
Angles & $k_\Theta$ (kcal/mol/rad$^2$) & $\Theta_\mathrm{min}$ ($^\circ$) \\
F-S-F & $73.461$ & $90.0$ \\
\hline 
Electrostatic Potential & \\
polarizable MDCM\cite{devereux:2020mdcm} * & \multicolumn{2}{c}{\url{https://github.com/MMunibas/N2O_SF6_Xe/}} \\
 & \multicolumn{2}{c}{in, e.g., \url{N2O_Xe/source/n2o_xe.dcm}} \\
 & \multicolumn{2}{c}{or \url{N2O_SF6/source/n2o_sf6.dcm}} \\
\hline 
Non-bonded & $\epsilon_i$ (kcal/mol)& $R_{\mathrm{min},i}/2$ (\AA) \\
S & $0.3257$ & $1.822$ \\
F & $0.0541$ & $1.659$ \\
\hline \hline
\textbf{Xe} & \multicolumn{2}{c}{Aziz and coworker\cite{aziz:1986}} \\
\hline 
Electrostatic Potential & $q$ (e) & $\alpha$ (1/\AA$^3$) \\
polarizable MDCM\cite{devereux:2020mdcm} ** & $0.0$ & $2.964$ \\
\hline 
Non-bonded & $\epsilon_i$ (kcal/mol)& $R_{\mathrm{min},i}/2$ (\AA) \\
Xe & $0.5610$ & $2.181$ \\
\hline \hline
\multicolumn{3}{l}{* Contributing only to N$_2$O-SF$_6$ interaction} \\
\multicolumn{3}{l}{** Contributing effectively only to N$_2$O-Xe interaction}
\end{tabular}
\end{table}

\clearpage

\section{Local Solvent Reorganization Lifetime}
The local solvent reorganization lifetime can be used as a proxy to
determine the critical concentration (density) at which transition so
a SCF occurs.\cite{tucker:2000a, tucker:2000b}

\begin{figure}[htb!]
\begin{center}
\includegraphics[width=0.80\textwidth]{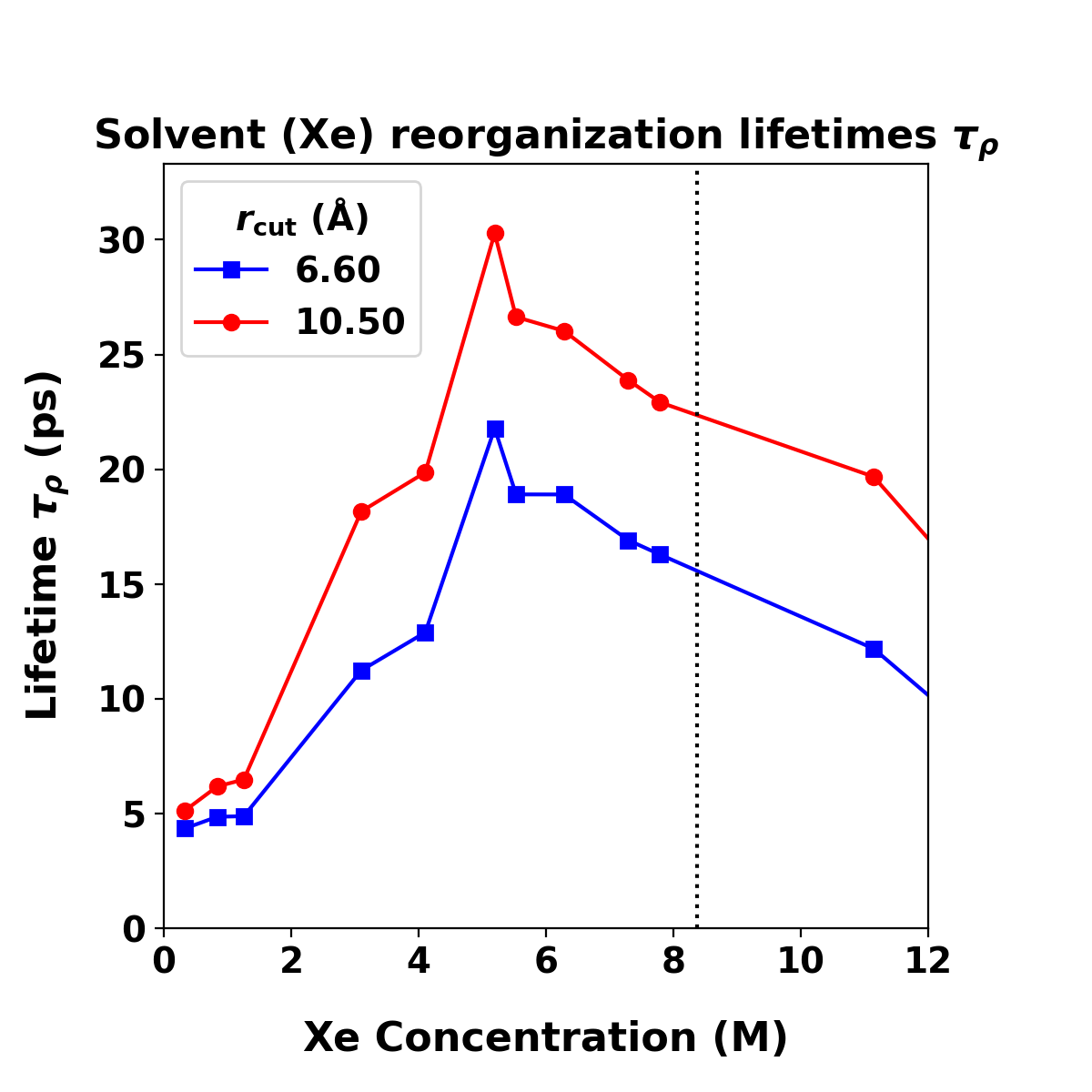}
\caption{Local solvent reorganization lifetime $\tau_\rho$ from MD
  simulations of pure Xenon (600 atoms) at $T = 291.2$\,K. The
  lifetime is computed following the work by Tucker {\it et
    al.}\cite{tucker:2000a,tucker:2000b} for two cutoff radii $r_{\rm
    cut} = {6.6, 10.5}$\,\AA\/ for the first and second solvation
  shells, respectively. These radii correspond to the local minima of
  the Xe--Xe radial distribution function, see Figure
  \ref{sifig:4}A. The peak in $\tau_\rho$ as a function of
  concentration at $c({\rm Xe}) = 5.19$\,M compares with an
  experimentally determined critical concentration of $c({\rm Xe}) =
  8.45$\,M (dotted vertical line).\cite{haynes:2014crc} This
  difference arises due to the fact that the parametrization of the
  Xe--Xe interaction potential was carried out exclusively based on
  reference data from the gas phase.\cite{aziz:1986}}
\label{sifig:1}
\end{center}
\end{figure}

\begin{figure}[htb!]
\begin{center}
\includegraphics[width=0.90\textwidth]{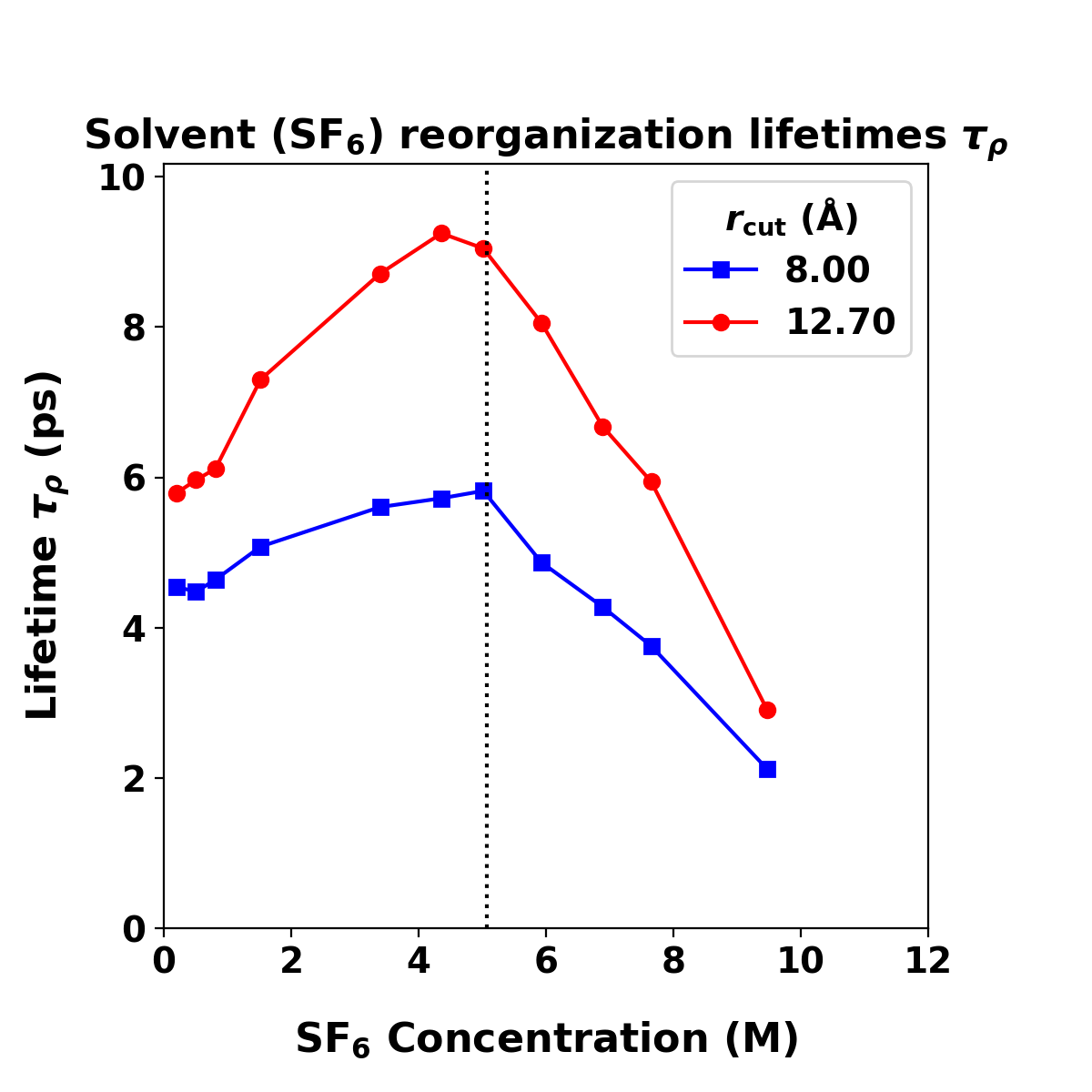}
\caption{Local solvent reorganization lifetime $\tau_\rho$ from MD
  simulations of pure SF$_6$ (343 molecules) at $T = 321.9$\,K.  The
  lifetime is computed following the work by Tucker {\it et
    al.}\cite{tucker:2000a,tucker:2000b} for two cutoff radii $r_{\rm
    cut}={8.0, 12.7}$\,\AA\/ of the first and second solvation shells,
  respectively. These radii correspond to the local minima of the
  SF$_6$--SF$_6$ radial distribution function, see Figure
  \ref{sifig:4}B. The peak in $\tau_\rho$ for
  $r_\mathrm{cut}=8.0$ at $c({\rm SF_6}) = 5.02$\,M agrees with the
  experimentally observed\cite{haynes:2014crc} critical concentration
  for SF$_6$ at $c({\rm SF_6}) = 5.06$\,M at $T = 318.76$\,K. }
\label{sifig:2}
\end{center}
\end{figure}

\clearpage

\section{Radial Distribution Functions}

\begin{figure}[htb]
\begin{center}
\includegraphics[width=0.90\textwidth]{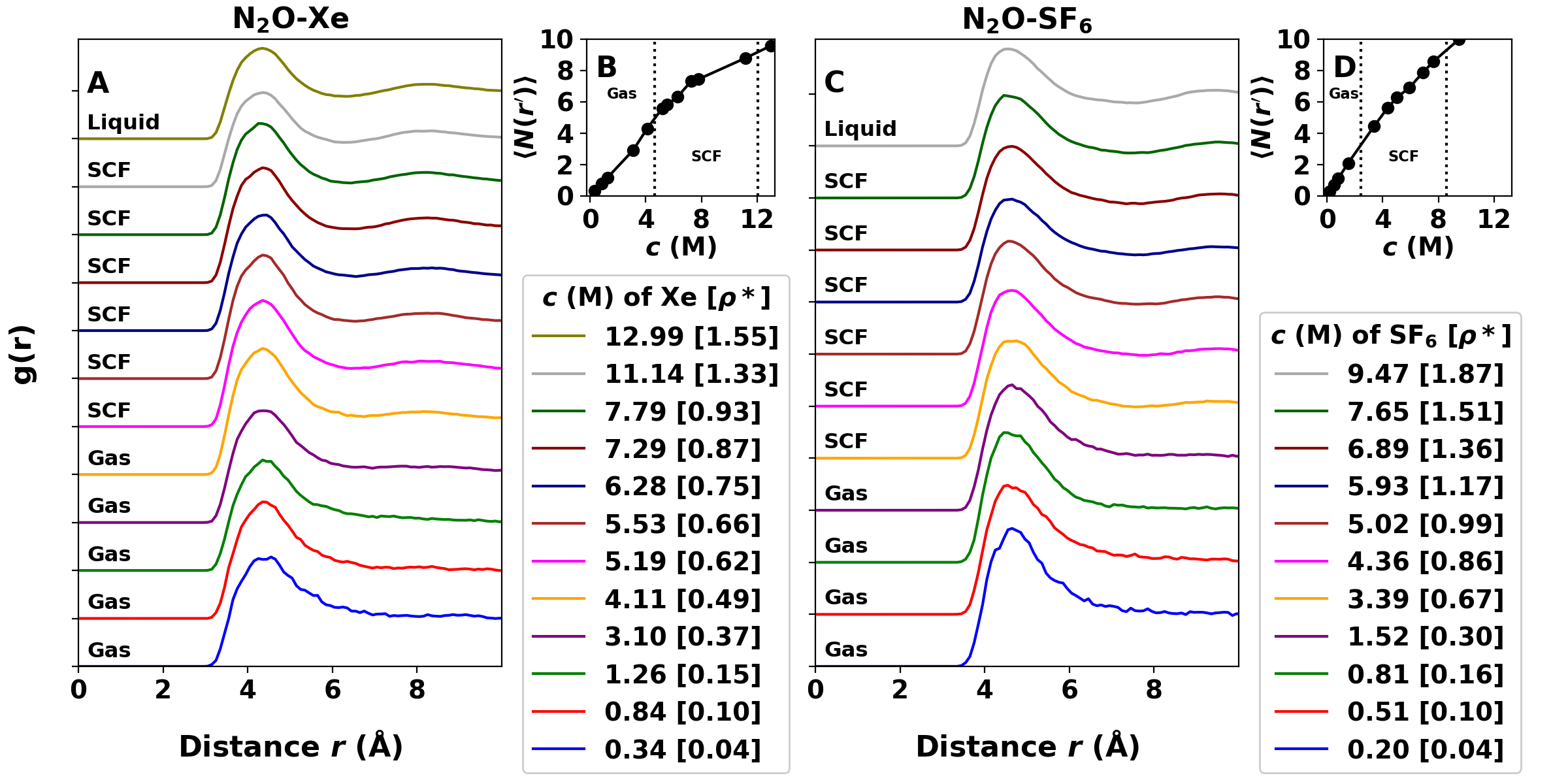}
\caption{Radial distribution function $g(r)$ between N$_2$O solutes
  center of mass and (A) the Xe atom and (C) the sulfur atom of SF$_6$
  at different solvent concentrations. Panels B and D show the
  coordination number of the Xe and SF$_6$ solvent compounds within
  the first solvation shell around N$_2$O of about (B) $r' =
  6.5$\,\AA\/ and (D) $r' = 7.5$\,\AA.  The vertical dotted lines mark
  the approximate concentrations of the solvent phase transition.
  Convergence of $g(r)$ for the gas phase systems (lower solvent
  density) is not completely achieved due to the small number of
  solvent molecules around a single N$_2$O solute. The gas phase
  $g(r)$ for $c = 0.34$ M in panel A is generated from an order of
  magnitude fewer distances within 10 \AA\/ compared with that for
  $c = 5.19$ M.}
  \label{sifig:3}
\end{center}
\end{figure}

\begin{figure}[htb]
\begin{center}
\includegraphics[width=0.90\textwidth]{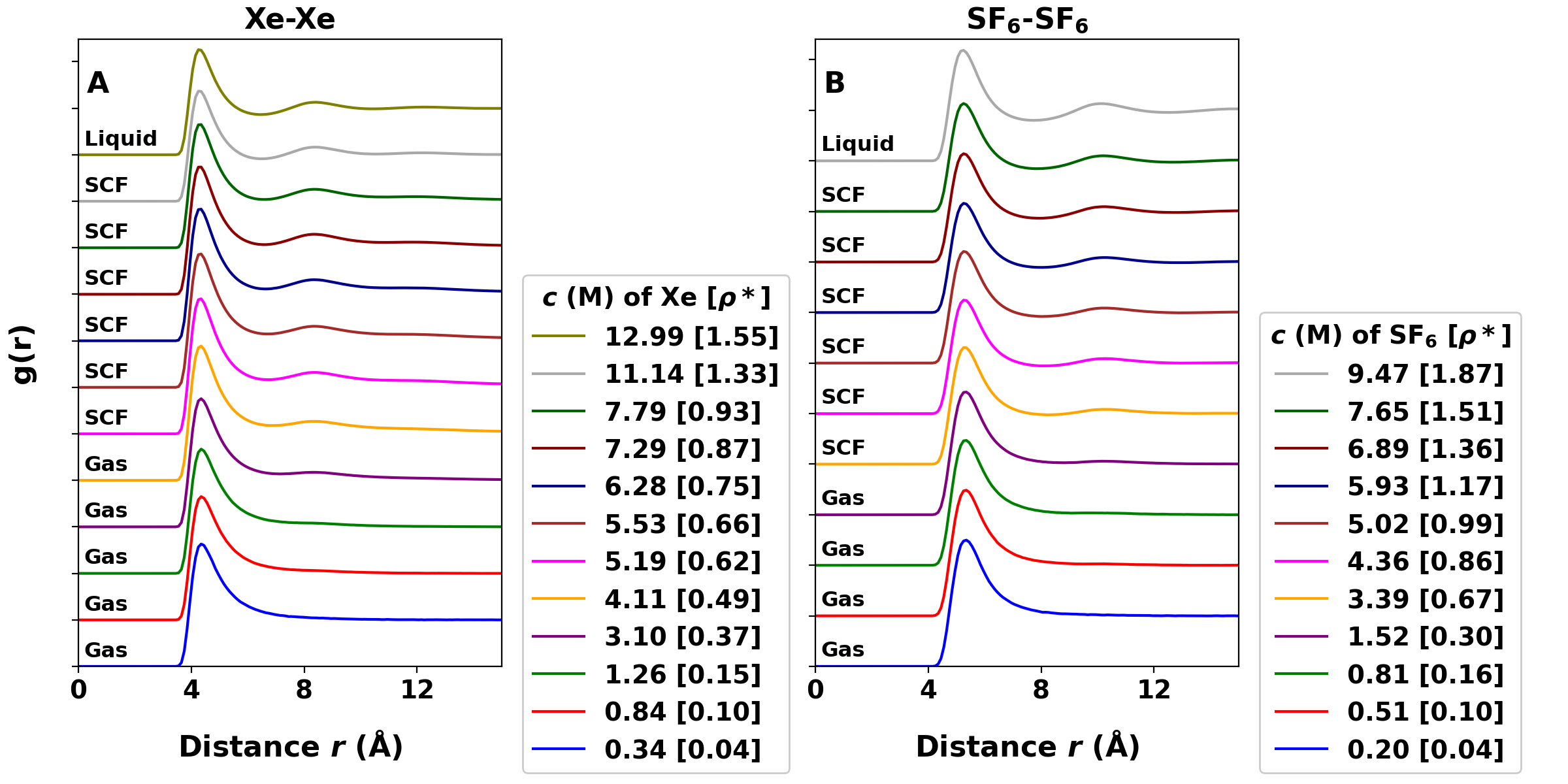}
\caption{Solvent-solvent radial distribution functions $g_{\rm SS}(r)$
  for (A) Xe and (B) SF$_6$ with the sulfur atom as the reference at
  different solvent concentrations.}
  \label{sifig:4}
\end{center}
\end{figure}

\clearpage

\section{Infrared Spectra}

\begin{figure}[htb!]
\begin{center}
\includegraphics[width=0.80\textwidth]{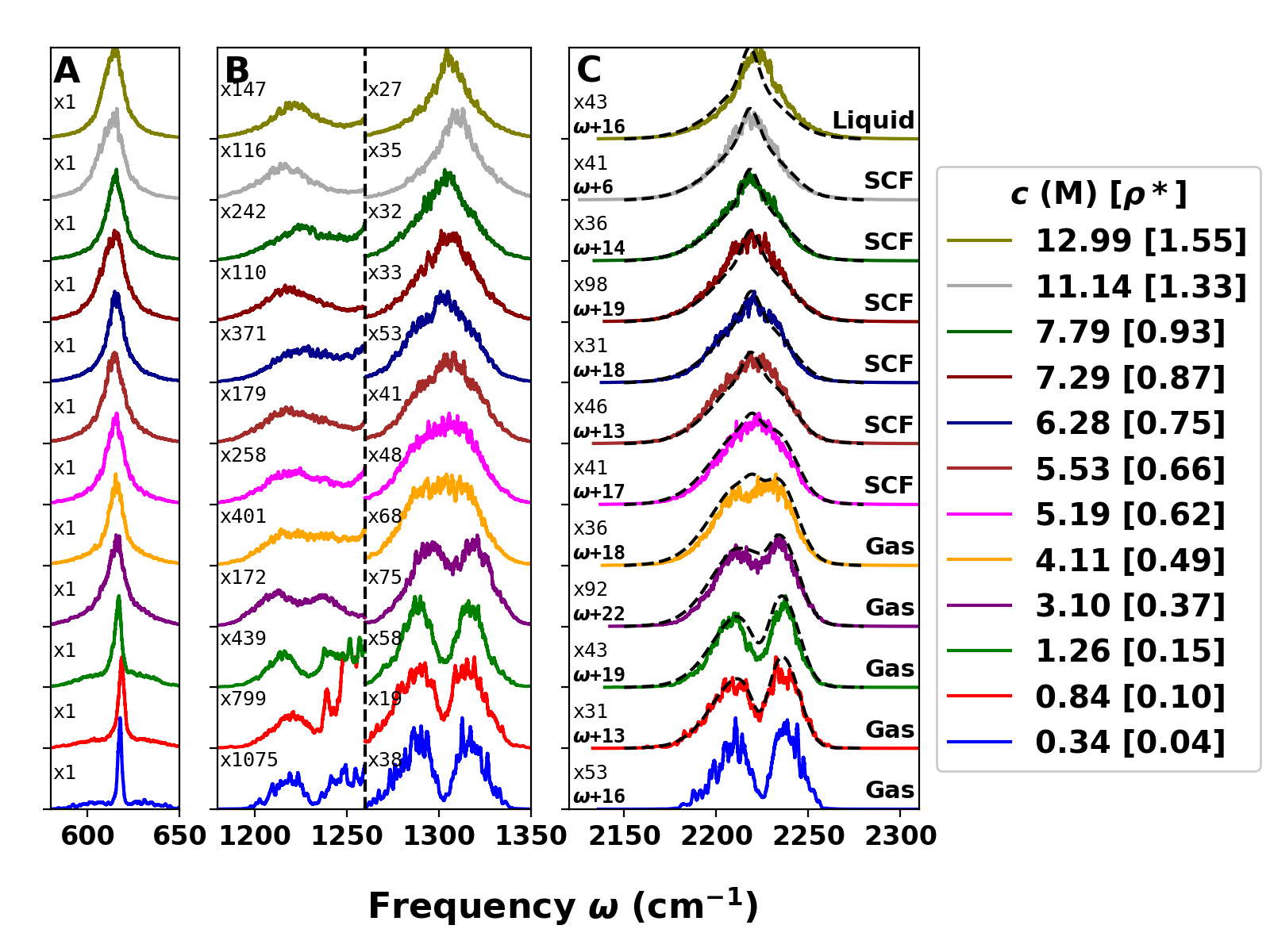}
\caption{IR spectra of N$_2$O in xenon at different solvent
  concentrations frequency ranges of (A) 580-650\,cm$^{-1}$, (B)
  1180-1350\,cm$^{-1}$ and (C) 2120-2310\,cm$^{-1}$ at $291.2$\,K.
  The amplitudes are scaled by the factor indicated at the center left
  for each range and density.  The amplitude of the line shape in
  Panel B are separately scaled for hot band and symmetric stretch on
  the left and right of the black dashed line, respectively.  In the
  right column, the line shape frequency bands are shifted in
  frequency $\omega$ to maximize the overlap with the experimental IR
  signal (dashed black lines, at 291\,K for gas, SCF, and at 287\,K
  for liquid xenon) at the corresponding density for the N$_2$O
  asymmetric stretch vibration.  The frequency shift is also given in
  the bottom left corner, and an average shift is applied for
  densities without experimental reference spectra.}
\label{sifig:5}
\end{center}
\end{figure}

\begin{figure}[htb!]
\begin{center}
\includegraphics[width=0.80\textwidth]{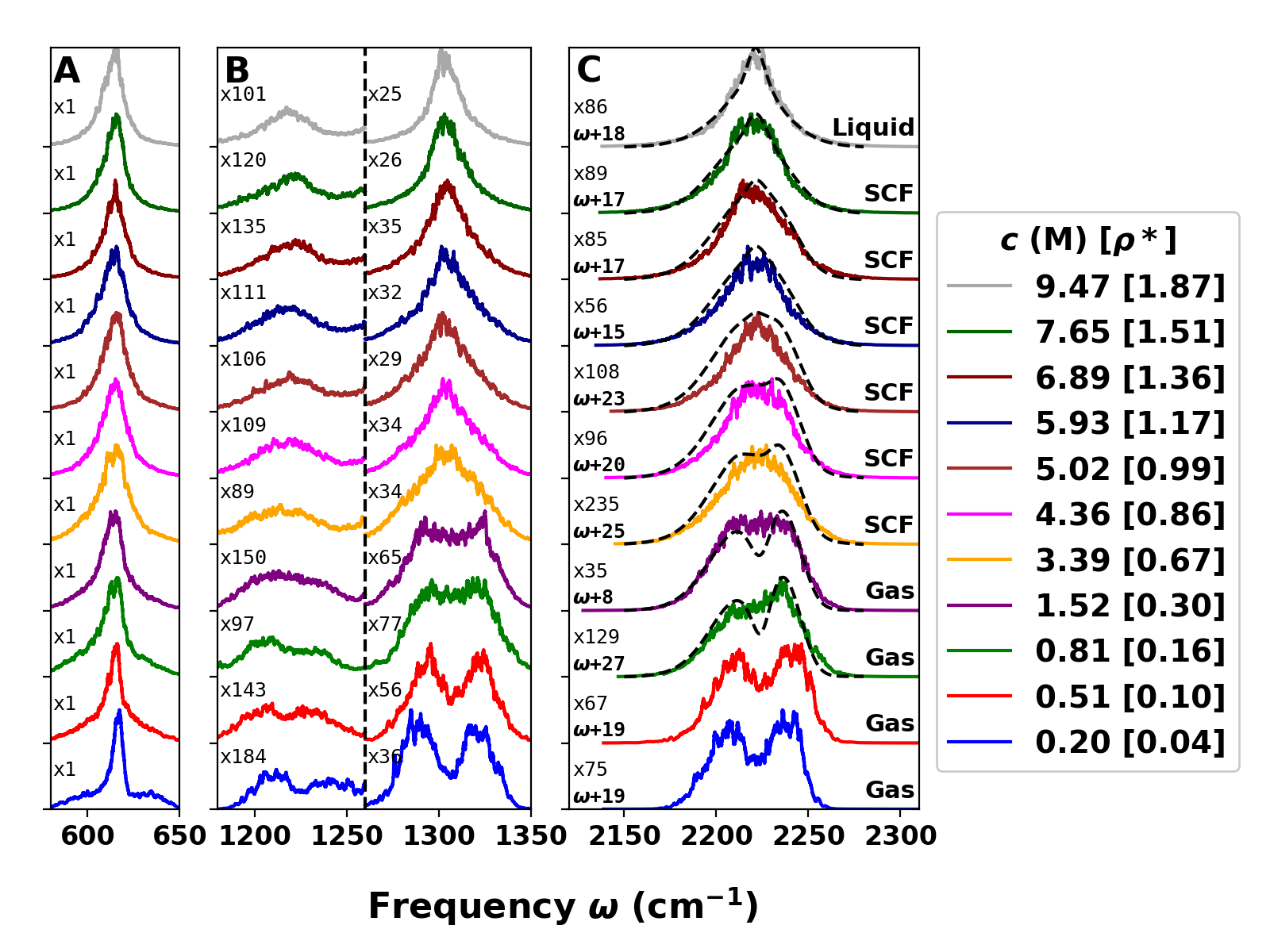}
\caption{IR spectra of N$_2$O in SF$_6$ at different solvent
  concentrations frequency ranges of (A) 580-650\,cm$^{-1}$, (B)
  1180-1350\,cm$^{-1}$ and (C) 2120-2310\,cm$^{-1}$ at $291.2$\,K.
  The amplitudes are scaled by the factor indicated at the center left
  for each range and density.  The amplitude of the line shape in
  Panel B are separately scaled for hot band and symmetric stretch on
  the left and right of the black dashed line, respectively.  In the
  right column, the line shape frequency bands are shifted in
  frequency $\omega$ to maximize the overlap with the experimental IR
  signal (dashed black lines, at 322\,K for gas, SCF, and at 293\,K
  for liquid SF$_6$) at the corresponding density for the N$_2$O
  asymmetric stretch vibration.  The frequency shift is also given in
  the bottom left corner, and an average shift is applied for
  densities without experimental reference spectra.}
\label{sifig:6}
\end{center}
\end{figure}

\clearpage

\section{Frequency-Frequency Correlation Function}

\begin{figure}[htb!]
\begin{center}
\includegraphics[width=0.90\textwidth]{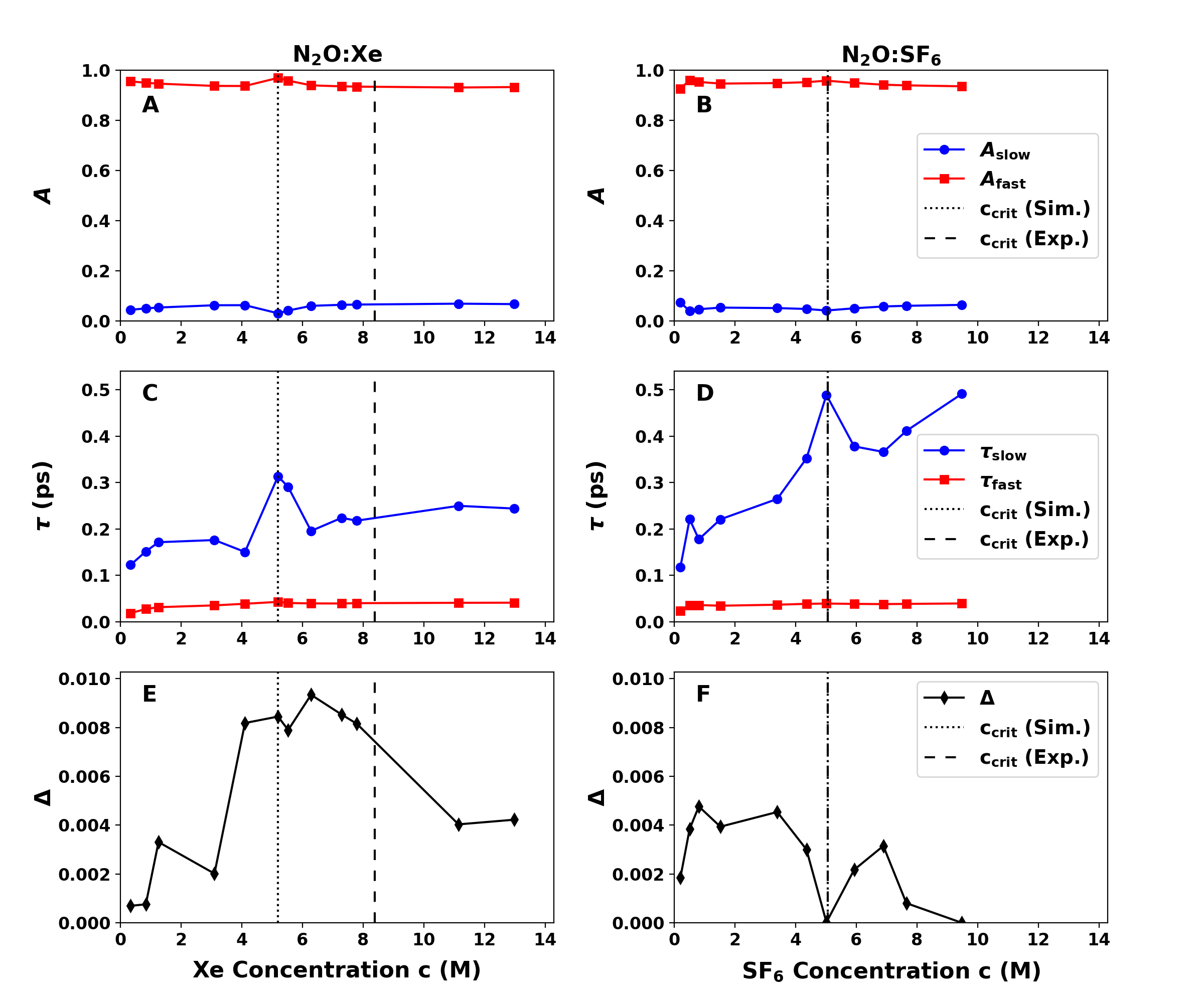}
\caption{Amplitudes ($a_1$ and $a_2$), decay times ($\tau_1$ and
  $\tau_2$), and asymptotic values ($\Delta$) for the bi-exponential
  fits of the normalized FFCF to the $\nu_\mathrm{as}$ time
  series from INMs for N$_2$O in Xe (A, C, E) and SF$_6$ (B, D, F)
  depending on solvent concentration.  The
  experimental\cite{haynes:2014crc} critical concentration
  $c_\mathrm{crit}$ and from simulations at $T_c$ of xenon and SF$_6$,
  are marked by the vertical dashed and dotted line, respectively.  It
  should be noted that the Xe--Xe interactions\cite{aziz:1986} were
  not optimized to reproduce the experimentally known $T_c$ for pure
  Xe whereas the model for SF$_6$ does.}
\label{sifig:7}
\end{center}
\end{figure}

\clearpage

\section{Instantaneous Normal Modes}

\begin{figure}[htb!]
\begin{center}§
\includegraphics[width=0.90\textwidth]{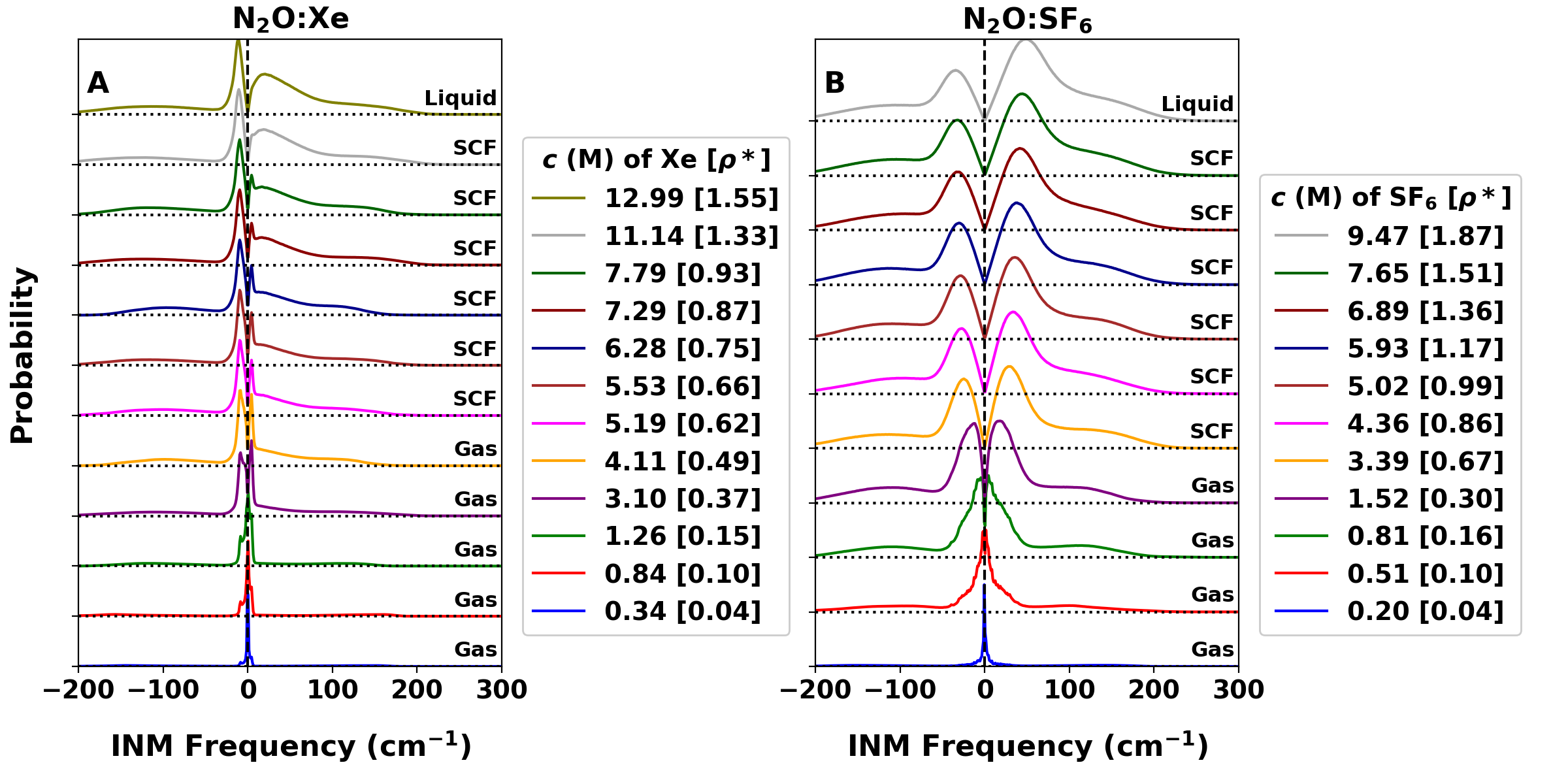}
\caption{Histogram of low-frequencies instantaneous normal modes
  frequencies of N$_2$O without prior geometry optimization in (A)
  xenon and (B) SF$_6$ at different solvent concentrations.  The
  vertical dashed line marks the zero position in the frequency range
  and the horizontal dotted line shows the zero probability for each
  solvent density to guide the eye.}
\label{sifig:8}
\end{center}
\end{figure}

\clearpage
\bibliography{ref_rotvib}